\documentclass[aps,prb,twocolumn,floats]{revtex4-2}
\usepackage{epsfig}
\usepackage{amsmath, amssymb}
\usepackage{graphicx}
\usepackage{soul}
\usepackage{subfigure}
\usepackage{color}

\newcommand{\blue}{\textcolor{blue}}

\begin{document}
\title{Topological Solitons in Su-Schrieffer-Heeger Chain with periodic hopping modulation, domain wall and disorder}
\author{Surajit Mandal, Satyaki Kar$^*$}
\affiliation{AKPC Mahavidyalaya, Bengai, West Bengal -712611, India}
\begin{abstract}
A chiral symmetric Su-Schrieffer-Heeger (SSH) chain features topological end states in one of its dimerized configurations. Those mid-gap zero energy states show interesting modifications upon a periodic tuning of the hopping modulations. Besides, more and more in-gap end modes appear at nonzero energies for further partitioning of the Brillouin zone (BZ) due to increased hopping periodicity. The new topological phases are identified with a detailed analysis of the topological invariants namely, winding number and Zak phases. The spectra and topology of these systems with periodically modulated hopping are studied also in the presence of a single static domain wall, separating two topologically inequivalent dimerized structures. {The domain wall causes additional in-gap modes in the spectrum as well as zero energy domain wall solitonic states for specific hopping periodicities.} We also study the effect of disorder, particularly the chirality breaking onsite ones, on the edge and domain wall states. {Other than the SSH type we also consider random, Rice-Mele or AI type disorder to do a comparative analysis of the evolution of chirality and zero energy states as the strength of disorder and hopping periodicity is varied}. Our findings can add important feedback in utilizing topological phases in various fields including quantum computations while the results can be easily verified in a cold atom set up within optical lattices.	
\end{abstract}
\maketitle
\section{Introduction}\label{sec1}
Topology in condensed matter\cite{nayak,elliot,beenakker,wilczek,ren,mourik,leijnse} has suddenly become a very hot topic these days due to the renewed understanding and subsequent discoveries both in theoretical\cite{ashwin,loss,jdsau,fu,skar} and experimental\cite{mourik,tewari} fronts identifying the robustness of the topological protections and its possible implementation in various fields including quantum computations\cite{meyer,vidal,horodechi}.

Today we often talk of graphene whose discovery\cite{scotch-tape} using scotch tape method in 2004 turned out a milestone in the journey of topological condensed matter systems, for its Dirac like excitations at low energies, high mobilities and topological stability\cite{sdsarma}. From there on, physicists have moved on a lot on topological system synthesis/analysis via introducing Haldane model\cite{haldane}, Kane-Mele model\cite{km}, topological insulators\cite{topo} and then Weyl, Dirac semimetals\cite{skrev} and so forth. Graphene has a staggered single bond-double bond structure (considering single resonance structure\cite{raji}) in a hexagonal lattice. But before it became popular, scientists were intrigued by the 1D conducting polymers involving the same staggered bonding structures. That's when the $(CH)_n$ Polyacetylene chain\cite{polyene} became important to the science community for its topological behavior, solitonic excitations, and domain wall structures\cite{ssh,wall2,ssh1}.

A long chain of Polyacetylene has a pair of degenerate ground states for two different sets of dimerization\cite{ssh}. For a finite chain these two staggered arrangements are topologically distinct for the end site sees either a strong or a weak bond. In the topological regime, end modes appear as zero energy states (ZES) that peak at the boundaries and die out exponentially away from there. One can call those end solitons. In a half-filled system, their spectral weights are equally shared between unoccupied conduction and occupied valence bands. With electrons added to or removed from the half-filled system, the end modes occupy with fractionalized charges $\mp e/2$. Strictly speaking, they are bonding and antibonding small gap states and of mixed chirality. But linear combinations of them can be considered to obtain a pair of chiral ZES that are located in single ends of the chain\cite{yang}.

A Su-Schrieffer-Heeger (SSH) model that can correspond to a polyacetylene chain, is a 1D tight-binding model with staggered hopping modulation\cite{ssh} and it  can demonstrate charge fractionalizations\cite{meier,ziani}, the existence of zero energy end states and topological solitonic excitations\cite{ssh,ssh1}. Neutral solitons with moving domain walls having $S=\pm\frac{1}{2}$ are obtained as the excitations though charged solitons can also found for a doped $(CH)_n$ system\cite{wall2}. The SSH chain has chiral symmetry and the chain features unit cells comprising of two adjacent sites. The staggered nature of the hopping brings in this two sublattice structure in the chain and accordingly the chiral symmetry, here, is also familiar as sublattice symmetry. It makes the Bloch Hamiltonian off-diagonal which gives the simplest route to derive the topological winding number, as we also have elaborated in this paper.

In this respect it will not be ludicrous to add a discussion on exotic Majorana fermions, that are their own antiparticles which appear as quasiparicle excitations, called Majorana bound states (MBS) in some condensed matter systems with defects\cite{nayak,elliot,beenakker}. Our system of concern - the SSH model can be broken down to two independent Majorana hopping chains. Similarly in a spinless Kitaev chain with
$p$-wave pairing\cite{kitaev}, a pair of independent Majorana hopping chains appear and single Majorana zero modes (MZM) survive there at the ends. The unconventional superconducting pairing can nullify the effect of quantum zero point motion to pin the Majorana states at the zero energy. Due to their non abelian exchange statistics, they lead to decoherence-free quantum information processing\cite{meyer}. However for the SSH model, the pair of Majorana modes localized at each end turn them into electronic modes and we do not get any Majorana physics in this system\cite{cmntMF}. But with interaction added, Majorana modes can be incorporated in such systems.
\begin{figure}[ht]
  \begin{center}
    \vskip -.6 in
    \begin{picture}(100,100)
     \put(-60,0){
       \includegraphics[width=0.8\linewidth]{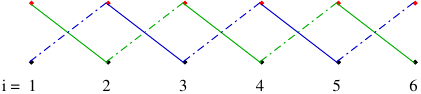}}
     \put(-60,30){(a)}
     \end{picture}
    \\\vskip -.5 in
    \begin{picture}(100,100)
      \put(-60,0){
        \includegraphics[width=0.8\linewidth]{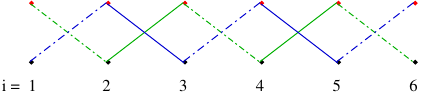}}
      \put(-60,30){(b)}
     \end{picture}
\end{center}
\caption{Both Kitaev chain (a) and SSH chain (b) can be broken down to two independent Majorana hopping models. When the weaker bond vanishes, end sites become disconnected from rest of the chain. While the Kitaev chain features 2 MBS localized at two ends, the SSH chain features 2 MBS at each end which turn them into electronic ZES.}
\label{fig0}
\end{figure}

The chiral symmetric SSH chain, with Hamiltonian as shown in Eq.(\ref{1}), features a topological phase for weak bonds at the boundaries, and midgap zero energy states (ZES) with fractional charge are obtained. Fig.\ref{fig0} demonstrates that while in a Kitaev chain end states are found to be Majorana modes, in the SSH model pair of Majorana modes at each end produces electronic ZES at two boundaries. Interesting becomes the case where ZES gets redistributed/modified due to a periodic variation of the hopping modulation that goes beyond simple staggered hopping of the SSH chain. A similar study has been initiated by one of the authors in Ref.\onlinecite{skar}. In this paper, we extend the calculations incorporating detailed analysis of the winding numbers to identify the topological/non-topological end modes of this periodically hopping modulated system. In addition here we also consider the effect of domain wall (DW) and disorders. {A static DW at the center of a SSH chain break it into two topologically inequivalent regime on its two sides and results in solitonic states in the system.} The consequence of having domain walls, in the SSH chain\cite{wall1} and other related models\cite{wall2,wall3,wall4,wall5} have been probed to some extent in the literature. {Here we find that the presence of periodic modulation of hopping amplitude causes additional in-gap modes to appear. Besides one of the end modes vanishes with new zero energy solitonic states appearing at the DW position for specific periodicities of hopping modulations.}
{ Furthermore, the addition of onsite disorder in our model leads to immediate disappearance of the chiral symmetry. We also consider random\cite{domain}, Rice-Mele (RM) and AI-type of disorder\cite{sang} for comparison which reveals that the disappearance of ZES with disorder strength is common in all of them though a DW state approaches zero energy for strong disorder of AI-type.}

The paper is organized as follows. In Sec.\ref{sec2} we provide the formulation of our SSH(like) model with periodic hopping modulation including also a detailed analysis of symmetry, topological invariant calculations as well as the features of spectra and end states. Next in Sec.\ref{sec4} we additionally introduce a single static domain wall at the center of a finite chain and study its response on the ZES and energy spectra. Sec.\ref{sec4a} describes the effect of onsite and hopping disorder in the periodically modulated chain with domain wall which affects the chirality of the system. Finally, we summarize our results in Sec.\ref{sec5} and discuss its novelty and possible future directions of work.

\section{Formulation}\label{sec2}
The Su-Schrieffer-Heeger (SSH) model \cite{ssh}, proposed in the context of polyacetylene is given by a one-dimensional tight-binding Hamiltonian with staggered nearest neighbor hopping [thus it shows a chiral (sub-lattice) symmetry]:
\begin{equation}\label{1}
  H_{SSH}=\sum_{i}^{L-1}(t+\delta_{i})c_{i}^{\dagger}c_{i+1}+h.c
\end{equation}
here $c_{i}^{\dagger}$, $c_{i}$ denotes electron creation and annihilation operator respectively and the periodic modulation in hopping strength (t) is obtained by $\delta_{i}=\Delta \cos[(i-1)\theta]$ in which $i=1,2,3,......, N$. In general, for $\theta=2\pi/n$, we get $\delta_{i+1}=\Delta \cos(\frac{2\pi i}{n})$ and the chain from a $n$ sublattice structure. Moreover, the system is represented by a $n\times n$ Hamiltonian matrix with $n$ number of eigenmodes. {The transformation $c_{i}\rightarrow(-1)^ic_{i}^\dagger$ gives
$H_{SSH}\rightarrow H_{SSH}$, implying the sublattice or chiral symmetry\cite{cmnt-chiral}.} For a model considering periodic boundary conditions (PBC), chiral symmetry also needs the total number of sites to be even.

With the introduction of Majorana operators as
\begin{equation}\label{1aa}
\gamma_{i,A}=\frac{1}{\sqrt{2}}(c_{i}^{\dagger}+c_{i})\;,\;  \gamma_{i,B}=\frac{i}{\sqrt{2}}(c_{i}^{\dagger}-c_{i}),
\end{equation}
the Majorana representation of Eq.(\ref{1}) becomes
\begin{equation}\label{1bb}
  H_{SSH}=\sum_{i}^{L-1}i(t+\delta_{i})[\gamma_{i,A}\gamma_{i+1,B}+\gamma_{i+1,A}\gamma_{i,B}].
\end{equation}
The case of $\theta=\pi$ corroborates the original SSH chain, which has been studied thoroughly. Here we depict the cases of few other $\theta$ values, that are commensurate to the finite chain.

\subsection{For $\theta=\frac{\pi}{2}$} 
Let us begin with a (finite) SSH(like) chain with an open boundary condition (OBC) for $\theta=\frac{\pi}{2}$:
\begin{eqnarray}\label{16}
H&=&\sum_{i}^{L-1}(t+\Delta)c_{i,A}^\dagger c_{i,B}+t c_{i,B}^\dagger c_{i,C}\nonumber\\&&+(t-\Delta)c_{i,C}^\dagger c_{i,D}+t c_{i,D}^\dagger c_{i+1,A}+h.c\ 
\end{eqnarray}
where $i$ denotes the unit cell. For PBC, we can consider the Fourier transform for the above Hamiltonian as
\begin{equation}\label{17}
c_{i,\alpha}=\frac{2}{\sqrt{L}}\sum_{{k\epsilon BZ}}e^{jki}\beta
\end{equation}
where $\alpha=\{A,B,C,D\}$ and $\beta=\{a_{k},b_{k},c_{k},d_{k}\}$ refer to the sublattice index and the corresponding Fourier modes respectively. Thus the Hamiltonian {[Eq.(\ref{16})] in terms of the spinor field $\psi_{k}=(a_{k},b_{k},c_{k},d_{k})^{T}$} takes the form of
\begin{eqnarray}\label{18}
H&=&\sum_{{k\epsilon BZ}}\psi_{k}^{\dagger}H_{k}\psi_{k}
\end{eqnarray} 
where the Bloch Hamiltonian 
\begin{equation}\label{18aa}
H_{k} = 
\begin{pmatrix}
0 & (t+\Delta) & 0 & t{e^{-4ik}} \\
(t+\Delta) & 0 & t & 0 \\
0 & t & 0 & (t-\Delta) \\
t{e^{4ik}} & 0 & (t-\Delta) & 0
\end{pmatrix}.
\end{equation}
The energy eigenvalues of the above matrix can be calculated as
\begin{equation}\label{19}
{\epsilon(k)=\pm\sqrt{2t^2+\Delta^2\pm t\sqrt{2t^2+6\Delta^2+2(t^2-\Delta^2) \cos4k}}}
\end{equation}
It essentially gives four bands in the energy spectrum. Notice that the energy gap closes for $\Delta=0$ at $k=0$ as well as for $(\Delta/t)^2=2$ at {$k=\pm\pi/4$}. Of these, the 1st case is a result of using periodic boundary implied in our Fourier construction but the system gaps out there in a finite chain. However the 2nd case indicates a gapless point even for the finite chain and correspond to the gap closing topological phase transition point. The bands cross linearly there (see Fig.\ref{fig2}) making the low energy modes behave as Dirac fermions\cite{dirac}.

\begin{figure}
   \vskip -.4 in
   \begin{picture}(100,100)
     \put(-70,0){
  \includegraphics[width=.33\linewidth,height=1 in]{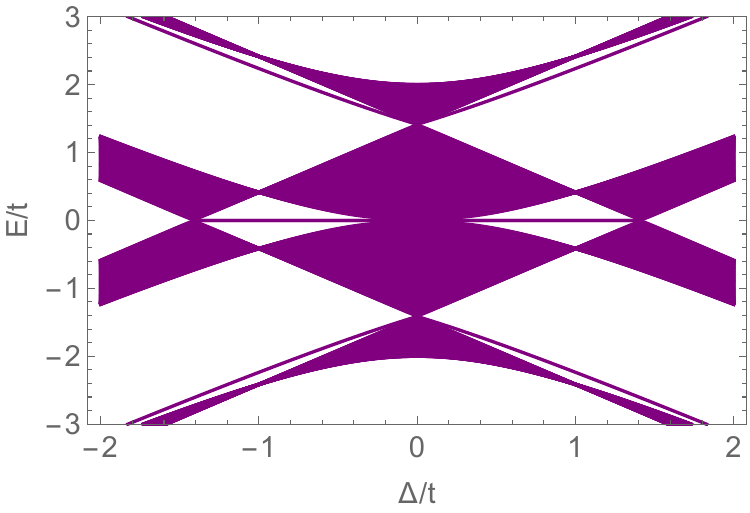}
  \includegraphics[width=.33\linewidth,height=1 in]{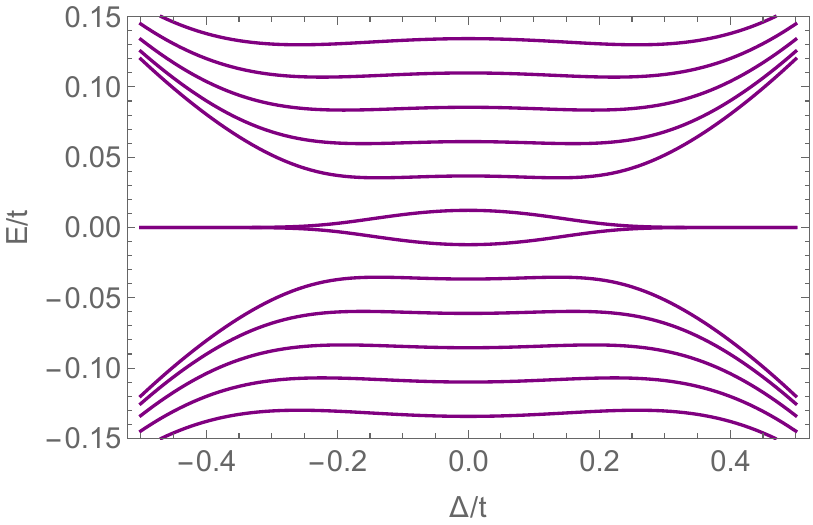}
  \includegraphics[width=.33\linewidth,height=1 in]{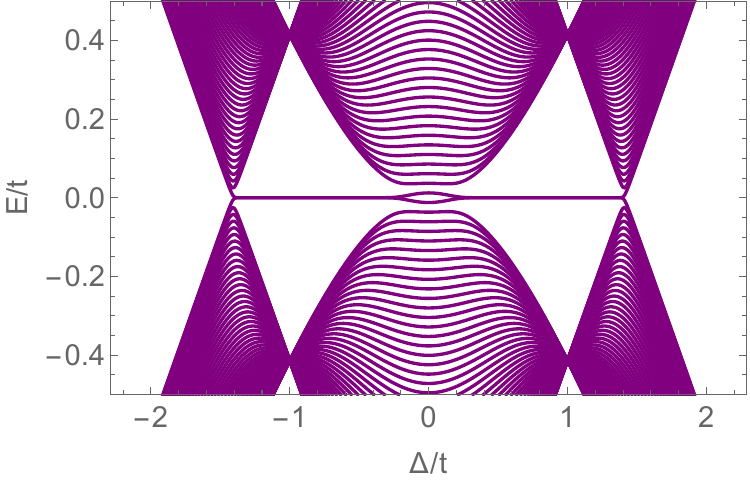}}
     \put(-45,52){(a)}
     \put(60,50){(b)}
      \put(156,48){(c)}
   \end{picture}\\
   \vskip -.2 in
   \begin{picture}(100,100)
     \put(-20,0){
   \includegraphics[width=.50\linewidth]{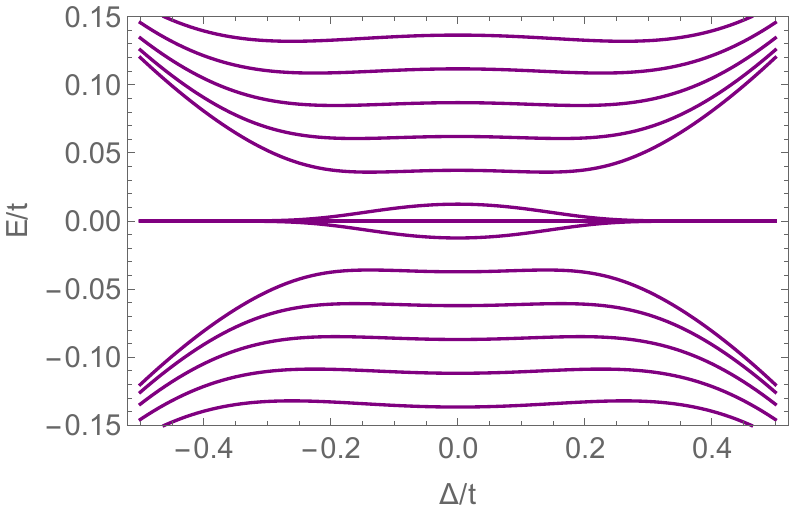}}
     \put(85,48){(d)}
   \end{picture}  
\caption{{(a)} Spectra of a SSH(like) chain {with $L=256$} considering OBC as a function of $\Delta/t$ for $\theta=\frac{\pi}{2}$. (b) and (c) shows the corresponding low energy states for $L=256$ with smaller and larger $\Delta/t$ respectively while (d) represents low energy spectra for an odd number of lattice sites with $L=255$.} 
\label{fig1}
\end{figure}

The numerical energy spectra for SSH(like) chain under OBC for $\theta=\pi/2$ is shown in Fig.\ref{fig1}. The figure indicates that energy spectra is symmetric (up-down) i.e. for every eigenstate with energy $E$, its chiral symmetric partner having energy $-E$ necessarily exists. Here, we get a 4 sublattice configuration yielding 2 in-gap non-zero energy states and 2 ZES. Zero energy modes with $L=256$ will no longer subsist for $-0.3\lesssim\Delta/t\lesssim 0.3$ and for $\Delta/t\gtrsim 1.4$ or $\Delta/t\gtrsim -1.4$ [see e.g. Figs.\ref{fig1}(b)-(c)].

For an odd number of lattice sites, one gets odd number of zero energy states. We examined that in a chain with odd number of sites, there are $L~modulo~4$ number of extra ZES that are localized exactly at $Mod(L,4)$ end sites. For example for $L=255$, with spectra as shown in Fig.\ref{fig1}(d), there are 5 (for $|\Delta/t|\gtrsim 0.25$) or 3 (for $|\Delta/t|\lesssim 0.25$) zero energy modes ($i.e.$, 3 extra ZES) with the extra modes being localized at site no. 253, 254 and 255 (it could have been at site number 1, 2, 3 on the other side of the chain had the Hamiltonian matrix taken the three extra bonds on that side).

\begin{table}[htb]
\centering
\begin{tabular}{ |p{1cm}|p{1cm}|p{1cm}|p{1cm}|p{1cm}|p{1cm}|p{1cm}|  }
 \hline
 Class& T &C&S&d=1&d=2&d=3\\
 \hline
 A & 0&0&0&-&\textbf{Z}&-\\
 \hline
  AI & +1&0&0&-&-&-\\
 \hline
  AII & -1&0&0&-&$\textbf{Z}_{2}$&$\textbf{Z}_{2}$ \\
 \hline
D & 0&+1&0&$\textbf{Z}_{2}$&\textbf{Z}&- \\
 \hline
 C & 0&-1&0&-&\textbf{Z}&- \\
 \hline
 AIII & 0&0&1&\textbf{Z}&-&\textbf{Z} \\
 \hline
 BDI & +1&+1&1&\textbf{Z}&-&- \\
 \hline
 CI & +1&-1&1&-&-&\textbf{Z} \\
 \hline
 DIII & -1&+1&1&$\textbf{Z}_{2}$&$\textbf{Z}_{2}$&\textbf{Z} \\
 \hline
 CII & -1&-1&1&\textbf{Z}&-&$\textbf{Z}_{2}$ \\
 \hline
\end{tabular}
\caption{Periodic table of topological invariants. Here, T, C, and S denote time-reversal symmetry, particle-hole symmetry, and and chiral symmetry operator respectively. The value of the corresponding single particle operator indicates the presence of symmetry. The different dimensions $d$ can host only trivial phases, indicated by $-$, a theoretically infinite number of phases labeled by an integer $\textbf{Z}$ or two phases, labeled by $\textbf{Z}_{2}$. This table is based on Altland’s and Zirnbauer’s ten symmetry classes\cite{s9d,s1}.}
\label{table1}
\end{table}

\subsubsection{Symmetry and Winding Number}
Now we discuss a bit more on the symmetry of the SSH(like) chain for $\theta=\frac{\pi}{2}$. By employing the {chiral symmetric operator} $\hat S=\Gamma_{4} = I_{2}\otimes \sigma_{z}$ with $I_{2}$ as the $2\times2$ identity matrix and $\sigma$ indicating the Pauli matrices, one can check that $\{\Gamma_{4},H_{k}\}=0$.
Hence, this Hamiltonian has a chiral symmetry (of course, for a chain with even number of sites). Apart from this, this Hamiltonian also respects the time-reversal symmetry ($T$) as well as particle-hole symmetry ($C$) and falls in the BDI class universality in a 10-fold way classification (see Table \ref{table1}). {Here $T$ ($C$) commutes (anti-commutes) with the Hamiltonian $H_{k}$ which, in this spinless case, is given by $\hat{T}=I_{2}\otimes I_2\mathcal{K}$ ($\hat{C}=I_{2}\otimes\sigma_{z}\mathcal{K}$), $\mathcal{K}$ denoting the complex-conjugation). The relation $\hat S=\hat T\hat C$ is thus satisfied\cite{ludwig}.} Due to the chiral symmetry of Hamiltonian, the Bloch Hamiltonian [Eq.(\ref{18aa})] is expected to be off-diagonal.  We should add here that the presence of chiral symmetry does not necessarily ensure $T$ and $C$ symmetry as there is a symmetry class called AIII that respects chiral symmetry alone. In fact, if we consider the hopping amplitudes to be complex, instead of real, then time-reversal symmetry no longer persists and $H_{SSH}$ moves on to the AIII universality class\cite{comment}. The AIII class, in one dimension ($1D$), is always topologically nontrivial\cite{s1}, unlike the usual SSH chain (with $\theta=\pi$), and its topology is described by a topological winding number.

There are two important corollaries of this chiral symmetry for the eigenstates and energies of the Hamiltonian. Firstly, the energy spectrum of the SSH(like) chain is always symmetric (up-down symmetry) in nature. For each eigenstate {$\psi_{k}$} having energy $E$, there is always another eigenstate (its chiral partner, {$\Gamma_{4}\psi_{k}=\psi^{\prime}_{k}$}) with energy $-E$. Secondly, states with zero energy only occupy a single sub-lattice. When the total number of particles is odd, chiral symmetry infers that there will be an odd number of end modes having zero energy. Moreover, the energy of the in-gap end states can also be non-zero, and this symmetry implies that such states having non-zero energy should be even in number and symmetrically positioned about $E=0$.

The one-dimension chiral models are identified by a \textbf{Z} topological index/invariant - winding number $\mathcal{W}$ which is an integer (may be positive or negative) while in two dimensions the equivalent index is Chern number\cite{s9a}.  The winding number is a completely mathematical property defined for any closed and smooth curve and is defined as the number of rotations (or windings) of the winding vector (defined below) about the origin\cite{s10} as one sweeps through the first Brillouin zone (BZ). The method of calculating the winding number is not only restricted to two-band models. It is much more general and can further be defined for any multi-band Hamiltonian that obeys chiral symmetry\cite{ole}. It is not always easy to determine the eigenvectors from Eq.(\ref{18aa}) in its analytic form. But for chiral symmetric systems, the definition of $\mathcal{W}$ can also be pertinent with the block off-diagonalized form of the Hamiltonian in the sub-lattice basis\cite{s2,s3}.

Applying the unitary transformation, the Hamiltonian $H_{k}$ in Eq.(\ref{18aa}) can be converted into block off-diagonal form :
\begin{equation}\label{6}
H_{k} = U H_{k}U^{-1}=
\begin{pmatrix}
0 & V \\
V^{\dagger} & 0 
\end{pmatrix}
\end{equation}
where, $V(k)(V^{\dagger}(k))$ is $2\times 2$ square matrices defined on the upper (lower) off-diagonal block of $H_{k}$, read as
\begin{equation}\label{7}
V(k)=
\begin{pmatrix}
 (t+\Delta) & t{e^{-4ik}} \\
 t & (t-\Delta)
\end{pmatrix}
\end{equation}
and $U$ is the unitary matrix obtained by employing the chiral-basis as
\begin{equation}\label{8}
U=
\begin{pmatrix}
1 & 0 & 0 & 0 \\\
0 & 0 & 1 & 0\\\
0 & 1 & 0 & 0\\\
0 & 0 & 0 & 1
\end{pmatrix}
\end{equation}
Now, the calculation of the winding number can be given by \cite{s9a,ole}
\begin{equation}\label{9b}
\mathcal{W}=-\frac{i}{2\pi}\int_{BZ} \partial_{k}i\phi(k)dk~.
\end{equation}
with integration on the reduced BZ, {$k\in$ [$-\pi/4$,\ $\pi/4$]\cite{bid}. Here $\phi_{k}$ is the phase of the complex number ${\bf Det[V(k)]}=(t^2-\Delta^2)-t^2e^{-4ik}=R(k)e^{i \phi(k)}$. For a two-band model (see e.g. Refs.\cite{s10,s11}) one does not need to consider the determinant as $V(k)$ is just a number. However, for systems with 4 bands or more, this formula aims to give the topological index without further diagonalizing the Hamiltonian Eq.(\ref{18aa})\cite{ryu}. One can substitute $\ln {\bf Det}[V(k)]$ in place of $i\phi_{k}$ in Eq.(\ref{9b})\cite{s9a}. Now, the winding number Eq.(\ref{9b}), considering the determinant, takes the form of \cite{s1,s2,s1a,s3,s3a}}
\begin{equation}\label{9}
\mathcal{W}=\int_{BZ}\frac{dk}{2\pi i} \partial_{k}[\ln {\bf Det}[V(k)]]
\end{equation}

and the estimate comes out to be 
\begin{equation}\label{9a}
{\mathcal{W} = \Bigg\{\begin{matrix}
     1, & 0<\Delta^2/t^2<2;\\
    0, & \Delta^2/t^2>2;\\
    \text{undefined}, &\Delta/t=0
  \end{matrix}}
\end{equation}
{The complex variable ${\bf Det[V(k)]}$ is called the winding vector (it has a magnitude and phase angle in the complex plane as already mentioned)} as its number of winding about the origin as $k$ varies in {[$-\pi/4$, $\pi/4$]} \cite{bid,s6} determines the winding number.

Eq.(\ref{9a}) implies that, one can get $\mathcal{W}=1$ for small $\Delta/t\ne0$ indicating the non-trivial topological phases (NTP) there. But $\Delta/t=0$ indicates an undefined ${\mathcal W}$ and hence no gap closing Lifshitz quantum phase transition (QPT) point. On the other hand, ${\mathcal W}$ remains unity till $0<|\Delta/t|<\sqrt{2}$ and vanishes for $|\Delta/t|>\sqrt{2}$. Thus QPT occurs at $\Delta=\pm\sqrt{2}t$ in this case. We should add here that interchanging the Majorana operators $\gamma_{i, A}$, $\gamma_{i, B}$ in Hamiltonian Eq.(\ref{1bb}) has no significant effect (other than an overall sign change) since our considered model is not including the next-nearest-neighbor hopping amplitudes and chemical potentials. Following the above procedure, we attain 
\begin{equation}\label{10}
V^{\prime}(k)=
\begin{pmatrix}
 (t+\Delta) & t \\
 t{e^{4ik}} & (t-\Delta)
\end{pmatrix}
\end{equation}
{causing merely a sign change in the winding number: $\mathcal{W}\rightarrow-\mathcal{W}$.}

It is worth mentioning that {the sign changes in winding number can occur} by relabelling sublattices {$\{A,B,C,D\} \Leftrightarrow \{D,C,B,A\}$} (or, interchanging the Majorana operators). This can also be visualized as a ‘choice’ of chiral symmetry operators. In the former case it is $\hat{S} = \Gamma_{4}$ acting in the sublattice basis while in the latter it is $\hat{S} =- \Gamma_{4}$. Moreover, the sign of the winding number is related to the winding direction of winding vector {\bf Det[V(k)]} which in turn leads to the type of (unpaired) Majorana fermion residing at the boundary. This change of sign makes no difference when one considers a single chain, however for multiple chains coupled to each other, such liberty no more persists as the relative sign of the individual chains becomes a relevant quantity of the system topology and Hamiltonian no more remains block off-diagonal. Detailed illustrations of the same can be found in Ref.\cite{s9a}.

To change the winding number from one value to another, we are required to either close the energy gap or disturb the chiral symmetry of the system, which is the pivotal symmetry of the Hamiltonian. Moreover, breaking or preserving the chiral symmetry after the implication of on-site and hopping disorder, respectively, are discussed in section \ref{sec4a}.

\begin{figure}
   \vskip -.4 in
   \begin{picture}(100,100)
     \put(-70,0){
  \includegraphics[width=.33\linewidth]{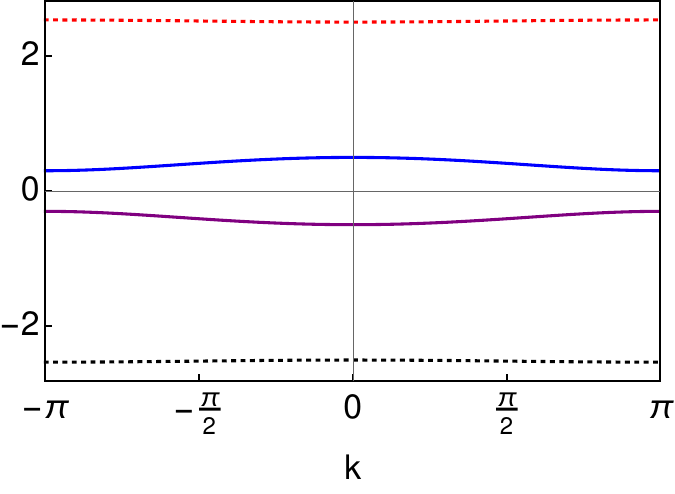}
  \includegraphics[width=.33\linewidth]{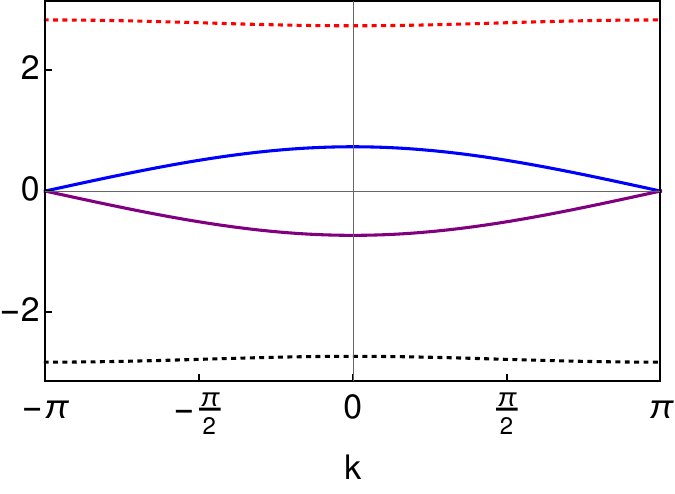}
   \includegraphics[width=.33\linewidth]{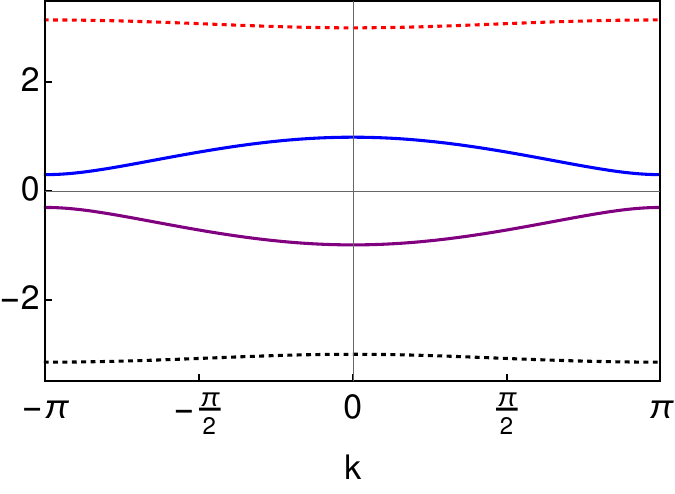}}
     \put(-45,48){(a)}
     \put(35,48){(b)}
     \put(120,48){(c)}
    \end{picture}\\
   \vskip .4 in
   \begin{picture}(100,100)
     \put(-30,0){
   \includegraphics[width=.6\linewidth]{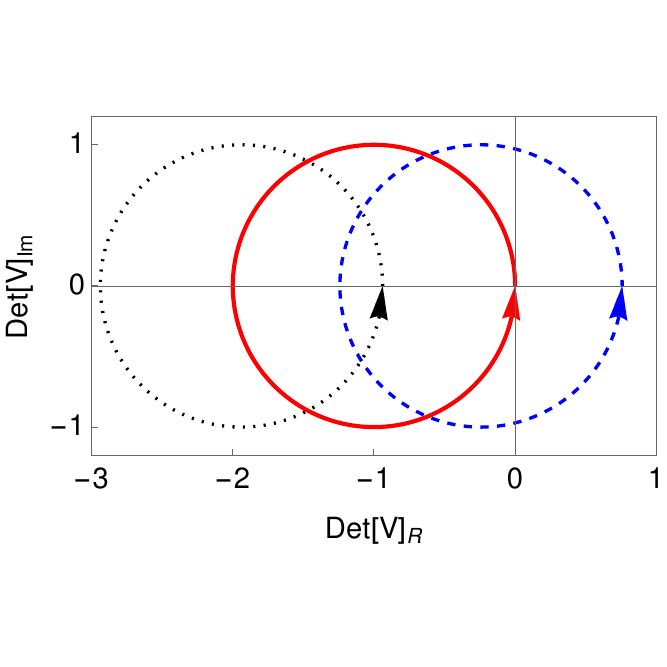}}
     \put(1,112){(d)}
   \end{picture}  
  \vskip -0.4 in
\caption{The dispersion plot corresponding to $\theta=\frac{\pi}{2}$ and $\Delta/t=$ (a) $\sqrt{2}-0.3$, (b) $\sqrt{2}$ and (c) $\sqrt{2}+0.3$ respectively. The bottom panel shows the winding of Det(V) as $k$ is varied through the {reduced} BZ, for these three cases in terms of dashed, solid and dotted circles respectively.} 
\label{fig2}
\end{figure}

The dispersion relation calculated from the bulk Hamiltonian Eq.(\ref{18aa}) is presented in the top panel of Fig.\ref{fig2}. Notice that the plot indicates asymmetry of the dispersion as $\Delta$ is varied about its critical value given by $(\Delta/t)^2=2$ - the topological phase transition points. 

In spirit of Eq.(\ref{9b}), one obtain
\begin{equation}\label{13}
\mathcal{W}=\frac{1}{2\pi}[\phi(\pi/4)-\phi(-\pi/4)]=\frac{1}{2\pi}\Delta\phi
\end{equation}
So if {\bf Det[V(k)]} takes $m$ revolutions with $\Delta\phi$ = $2\pi m$,  Eq.(\ref{13}) leads to $\mathcal{W}=m$.
The eigenstates are parameterized by $\phi=\tan^{-1}(\frac{Im[Det(V)]}{Real[Det(V)]})$. Therefore, the direction of the winding vector {\bf Det[V(k)]} in the complex space represents an eigenstate and the magnitude of the same estimates its eigenvalue. The trajectory of {\bf Det[V(k)]} parameterizes the evolution of the state as $k$ is varied from {$-\pi/4\rightarrow \pi/4$. The trajectories for the asymmetric dispersion about the critical value (mentioned above) }may or may not encircle the origin indicating non-trivial or trivial topological classes respectively.

The demonstration of the topological winding number for the 1D SSH chain (two sub-lattices) can be found in Ref.\cite{s10}.
In this context, the parametric plot of the winding number in the complex plane is depicted in Fig.\ref{fig2}. The arrowhead stipulates the direction of movement of {\bf Det[V(k)]} as $k$ is varied from {$-\pi/4\rightarrow\pi/4$}. The sense of rotation is in the counterclockwise direction for each of the closed contours. The contour for $\Delta/t=\sqrt{2}-0.3$ encloses the origin making $\mathcal{W}=1$ indicating the NTP while that for $\Delta/t=\sqrt{2}+0.3$ does not enclose the origin and indicates a trivial topological phase (TTP) with $\mathcal{W}=0$. The contour for $\Delta/t=\sqrt{2}$ touches the origin and indicates the gap closing phase transition point\cite{s5a}. This justifies our theoretical findings very well.

{In {\bf Appendix A}, we calculate the winding number in terms of poles and zeros and the Berry phase for this model is studied in {\bf Appendix B} [see Eq.(\ref{berrya})].} Notice that the 1D Polarization is the Berry phase of the occupied Bloch states. Electrostatics shows us that polarization is related to bound charges and in the SSH chain, the Berry phase thus estimates the charges of the end modes that comes out to be $\pm e/2$\cite{kane}.

As a side mark, we mention here that the topological nature of phase can also be determined based on the availability of the zero eigenvalues of the local in-gap Greens function as studied in Ref.\cite{slager}.

\subsubsection{Bulk-Boundary Correspondence}
As we mentioned, the bulk-boundary correspondence refers to the connection between edge/surface states of a finite system to the difference in bulk topological invariants across the QPT\cite{s10,bulk1,bulk2,bulk3,bulk4}. The topological behavior is associated with the existence of edge states on the boundary of open systems\cite{s10} which in turn is complemented by a nonzero Berry phase within the bulk (and vice-versa). The edge modes peak at the boundary and decays exponentially towards the bulk. Generally, bulk-boundary correspondence is present in most of the Hermitian systems. However, a more generalized understanding of the same is required for more complicated Hermitian and some non-Hermitian systems \cite{bulk5,bulk6,bulk7}.

For a system with chiral symmetry, one can find a chiral partner for each of the states. For the topological insulators, bulk-boundary relation refers to a one-to-one correspondence between the number of gapless zero-energy end-modes in a topologically non-trivial system and the topological invariant (winding number). For instance, the number of conducting edge states in a 2D quantum Hall system is the same as the Chern number that appears in the transverse conductivity\cite{s5a}. Unlike in the case of $\theta=\pi$, where end states are restricted in single sublattices, here for $\theta=\pi/2$ the end states are restricted in two sublattices such that no two consecutive sites see nonzero amplitude of them {In other words, amplitude of the end state wavefunctions vanishes in alternate sites (corresponding to two sublattices) while the rest of the sites (corresponding to the other two sublattices) alternately see positive and negative amplitudes also indicating a gradual decay of strength from the end of the chain towards the bulk.} The winding number is equivalent to the net number of end states: $N_{A}-N_{D}$, where $N_{A}$ and $N_{D}$ are the number of end states of sub-lattice $A$ and $D$ (that include the two end sites) respectively, on sublattice A on the left end (see pp. 16-17 of Ref.\cite{s10}). The winding number is estimated from the bulk Hamiltonian while the net number of end states can be obtained by looking at the low energy sector of the left end. For the trivial case, {$\Delta^2/t^2>2$}, both are 0 and for the topological case, {$0<\Delta^2/t^2<2$}, both are 1. This indicates that there is one end mode localized in sub-lattice $A$ and no end mode localized in sub-lattice $D$, observed from the left. The corresponding wave functions are shown to be highly located near the ends. Since each chiral pair is connected with opposite energies $\pm \epsilon(k)$, the single end mode must obey it as well resulting in $\epsilon(k)=0$. Interestingly, the wave functions of the zero energy end modes can be explicitly derived without solving the eigenvalue problems and further studied as in Ref.\cite{winding}.

Edge states in BDI class are time-reversal symmetric and there can be a phase factor of $\pm 1$. Thus, one may require the correct linear combination of degenerate edge states to notice this symmetry\cite{s9a}.

In the following, we will discuss the cases for two other commensurate $\theta$ values. To make the article precise, we only mention the distinctive features of the same.

\subsection{For $\theta=\pi/3$}
The chiral symmetry of an odd-$n$ SSH(like) model Eq.(\ref{1}) is obscured in momentum space as found in Ref.\cite{s3a}. Hence, the winding number is also not well defined for $\theta=2\pi/3$. Consequently, the chiral symmetry for this case does not give directly to the nontrivial band topology via the winding number. The possible way to make the chiral symmetry more diaphanous is assembling two neighboring unit cells of the SSH(like) model (for $\theta=2\pi/3$) simultaneously to constitute a six-band SSH model as documented in Ref.\cite{s3a,bulk7}. This six-band SSH model corresponds to $n=6$ or $\theta=\pi/3$.
{The corresponding Bloch Hamiltonian will be given by $6\times6$ matrix $H_k$ as shown in Eq.(\ref{32cc}) of {\bf Appendix C}.}

\begin{figure}
   \vskip -.4 in
   \begin{picture}(100,100)
     \put(-70,0){
  \includegraphics[width=.48\linewidth]{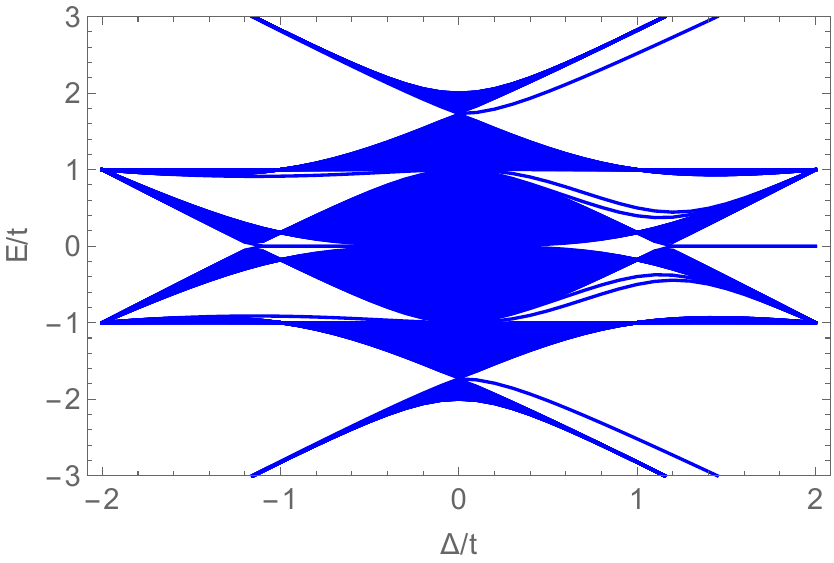}
  \includegraphics[width=.48\linewidth]{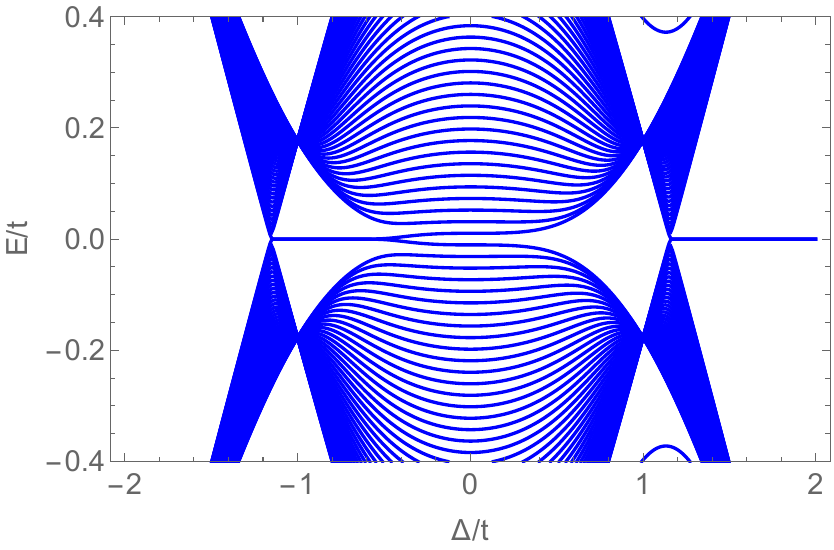}}
     \put(-45,60){(a)}
     \put(72,58){(b)}
   \end{picture}  
\caption{Spectra of a SSH(like) chain (considering OBC) as a function of $\Delta/t$ for $\theta=\frac{\pi}{3}$. (b) represents the same zoomed in at low energies. Both plots are for $L=300$.}
\label{fignew}
\end{figure}

The numerical spectra for this case is presented in Fig.\ref{fignew}. We see from the plot that energy spectra is symmetric i.e. for every eigenstate with energy $E$, its chiral symmetric partner having energy $-E$ necessarily exists. Here, similar to $\theta=\pi/2$, we get a 6 sublattice configuration and new gaps can be found within the spectrum. Interestingly we get 6 in-gap states in TTP while 2 in-gap states and 2 midgap ZES in NTP. Here, similar to $\theta=\pi/2$, ZES no longer exists for all values of $|\Delta/t|$ including $|\Delta/t|=0$, as shown in Fig.\ref{fignew}(b). The energy gap closes (and reopens) at $|\Delta/t|=0,~2/\sqrt{3}$ indicating three topological QPT points.

The above Hamiltonian has chiral symmetry and also respects time-reversal and particle-hole symmetry. Thus it belongs to the BDI class in the 10-fold way classification (see Table \ref{table1}). The chiral symmetry exhibits itself as $\Gamma H_{k} \Gamma =-H_{k}$ with $\Gamma_{ij}=(-1)^{i}\delta_{ij}$. The Hamiltonian $H_{k}$ in Eq.(\ref{32cc}) can be cast into block off-diagonal form as:
\begin{equation}\label{32dd}
H_{k} = 
 \begin{pmatrix}
0 & V \\
V^{\dagger} & 0 
\end{pmatrix}
\end{equation}
with 
\begin{equation}\label{32ee}
V=
\begin{pmatrix}
 (t+\Delta) & 0 & (t+\frac{\Delta}{2}){e^{-6ik}} \\
 (t+\frac{\Delta}{2}) & (t-\frac{\Delta}{2}) & 0 \\
 0 & (t-\Delta) & (t-\frac{\Delta}{2})\\
\end{pmatrix}
\end{equation}
Using the similar approach as shown for the $\theta=\pi/2$ case, one can obtain the winding number for the winding vector $\textbf{Det[V(k)]}$ to be

\begin{align}\label{32hh}
{\mathcal{W} = \Bigg\{\begin{matrix}
     1, & 0>\frac{\Delta}{t}>-2/\sqrt{3}~{\rm or}~\frac{\Delta}{t}>2/\sqrt{3}\\
    0, & 0<\frac{\Delta}{t}<2/\sqrt{3}~{\rm or}~\frac{\Delta}{t}<-2/\sqrt{3}.
  \end{matrix}}
\end{align}

Now in the parametric plot, similar to the case with $\theta=\pi/2$, the circular contour of winding vector ${\bf Det[V(k)]}$ for $t^3-3t\Delta^2/4+\Delta^3/4 > (<)~ t^3-3t\Delta^2/4-\Delta^3/4$ would not (would) encircle the origin suggesting a TTP (NTP) with winding number $\mathcal{W}=0~(1)$. And the contour should pass through the origin for $|\Delta/t|=0,~2/\sqrt{3}$ indicating a gap closing phase transition point.

\subsection{For $\theta=\pi/4$}

\begin{figure}
   \vskip -.4 in
   \begin{picture}(100,100)
     \put(-70,0){
  \includegraphics[width=.48\linewidth]{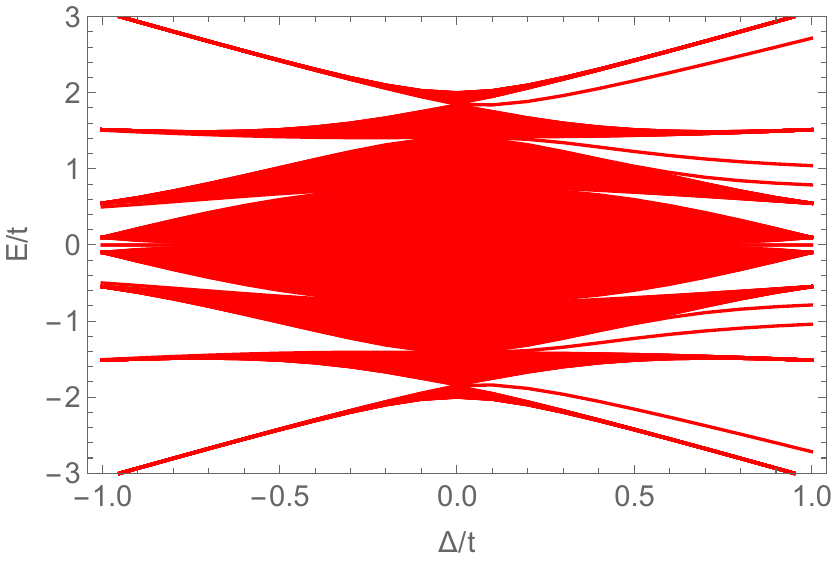}
  \includegraphics[width=.48\linewidth]{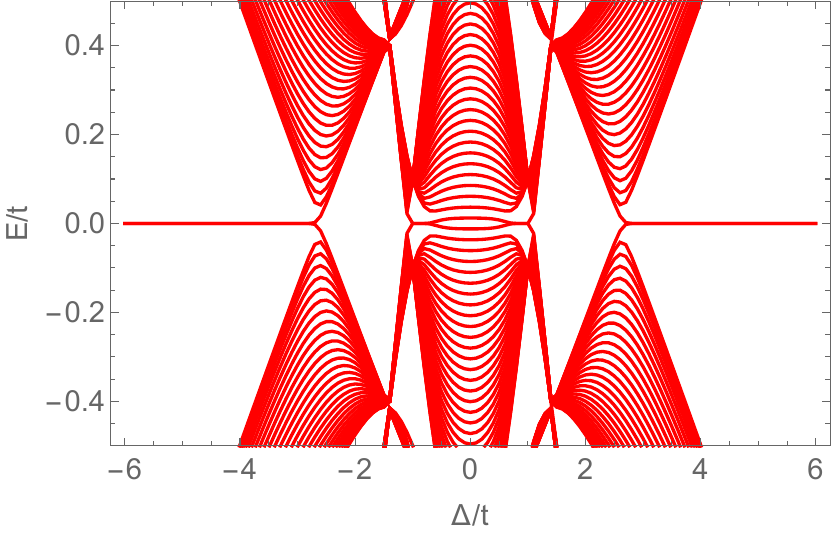}}
     \put(-45,63){(a)}
     \put(76,58){(b)}
   \end{picture}  
\caption{(a) Spectra of a SSH(like) chain with OBC as a function of $\Delta/t$ with $L=256$ for $\theta=\frac{\pi}{4}$. (b) represents the same zoomed in at low energies.}
\label{fig4a}
 \end{figure}

Now we consider a finite chain with $\theta=\frac{\pi}{4}$ for which one obtains a $8\times8$ Bloch Hamiltonian $H_k$ {[see  Eq.(\ref{32d}) of {\bf Appendix C}]}. The numerical spectra for this case is presented in Fig.\ref{fig4a}. We see from the plot that energy spectra is symmetric i.e. for every eigenstate with energy $E$, its chiral symmetric partner having energy $-E$ necessarily exists. The spectra confirm 8 in-gap states. Out of 8 in-gap states, 6 are found at non-zero energies while 2 are ZES. The existence of ZES with variation of $\Delta/t$ is shown in Fig.\ref{fig4a}(b). Spectrum shows the gap closing transitions to occur at $|\Delta/t|=\sqrt{2(2\pm\sqrt{2})}$ which are indeed the four topological QPT points in this case.

Just like in the previous cases, an unitary transformation from there leads to the block diagonalization:

\begin{equation}\label{32e}
H_{k} \rightarrow U H_{k}U^{-1}=
 \begin{pmatrix}
0 & V \\
V^{\dagger} & 0 
\end{pmatrix}
\end{equation}
with 
\begin{equation}\label{32f}
V=
\begin{pmatrix}
 (t+\Delta) & 0 & 0 & (t+\frac{\Delta}{\sqrt{2}}){e^{-8ik}} \\
 (t+\frac{\Delta}{\sqrt{2}}) & t & 0 & 0 \\
 0 & (t-\frac{\Delta}{\sqrt{2}}) & (t-\Delta) & 0 \\
 0 & 0 & (t-\frac{\Delta}{\sqrt{2}}) & t
\end{pmatrix}.
\end{equation}

This now helps in estimating the winding number and we obtain
\begin{equation}\label{32j}
{{\mathcal{W}} = \Bigg\{\begin{matrix}
     1, & 0<\frac{\Delta^2}{t^2}<2(2-\sqrt{2})~{\rm or}~ \frac{\Delta^2}{t^2}>2(2+\sqrt{2})\\
    0, & 2(2-\sqrt{2})<\Delta^2/t^2<2(2+\sqrt{2})\\
    \text{undefined}, &\Delta^2/t^2=2(2\pm\sqrt{2}).
  \end{matrix}}
\end{equation}

One can again look at the parametric plot of {\bf Det[V(k)]} in the complex plane for all Brillouin zone vectors and we find the circular {contour to enclose the origin for $\Delta^2/t^2 >(<)~ [2(2+(-)\sqrt{2})]$ signaling a NTP with winding number $\mathcal{W}=1$.}
\begin{figure}
   \vskip -.4 in
   \begin{picture}(100,100)
     \put(-70,0){
  \includegraphics[width=.48\linewidth]{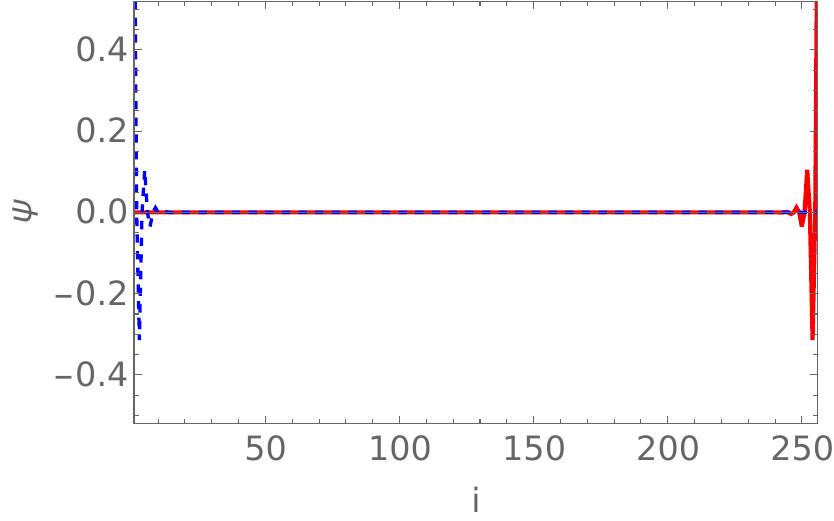}
  \includegraphics[width=.48\linewidth]{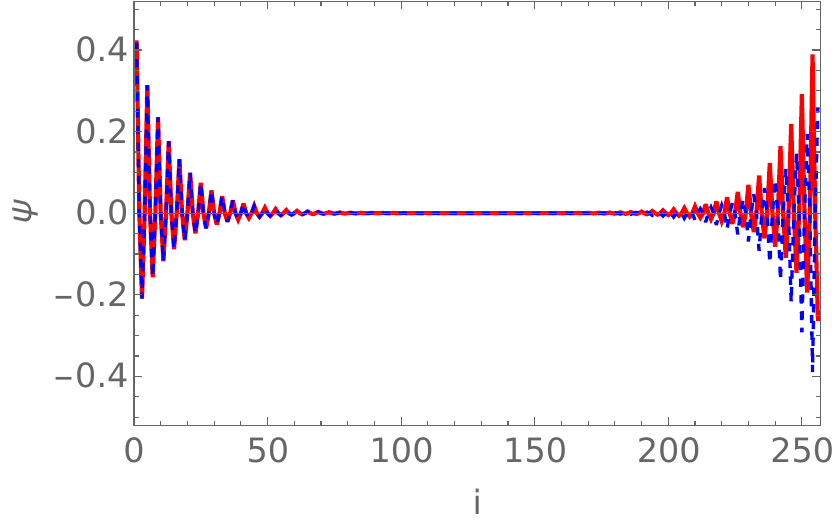}}
     \put(-10,60){(a)}
     \put(110,60){(b)}
   \end{picture}\\
   \vskip -.2 in
   \begin{picture}(100,100)
     \put(-70,0){
   \includegraphics[width=.48\linewidth]{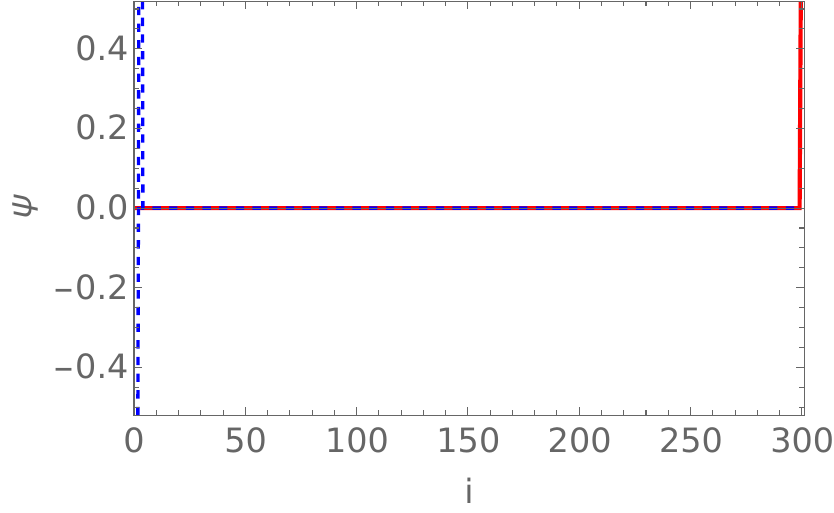}
   \includegraphics[width=.48\linewidth]{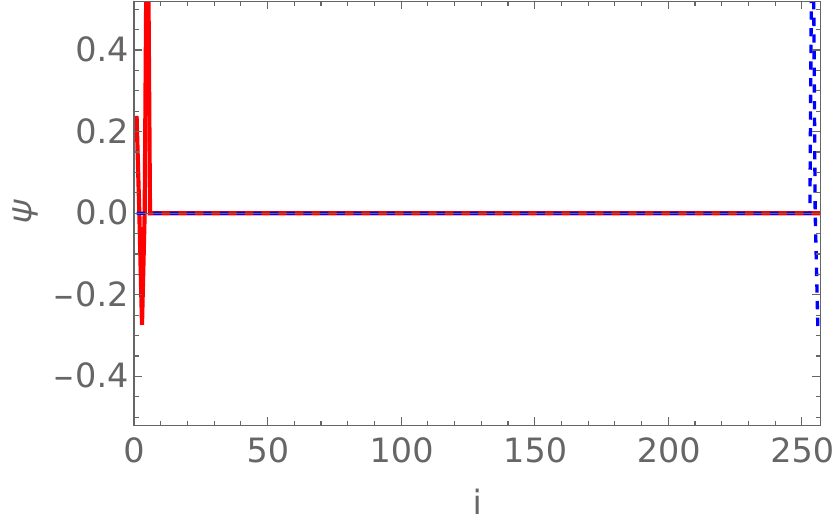}}
   \put(-10,60){(c)}
     \put(110,60){(d)}
   \end{picture}  \\
   \vskip 1 in
   \begin{picture}(100,100)
   \put(-70,0){
     \includegraphics[width=\linewidth]{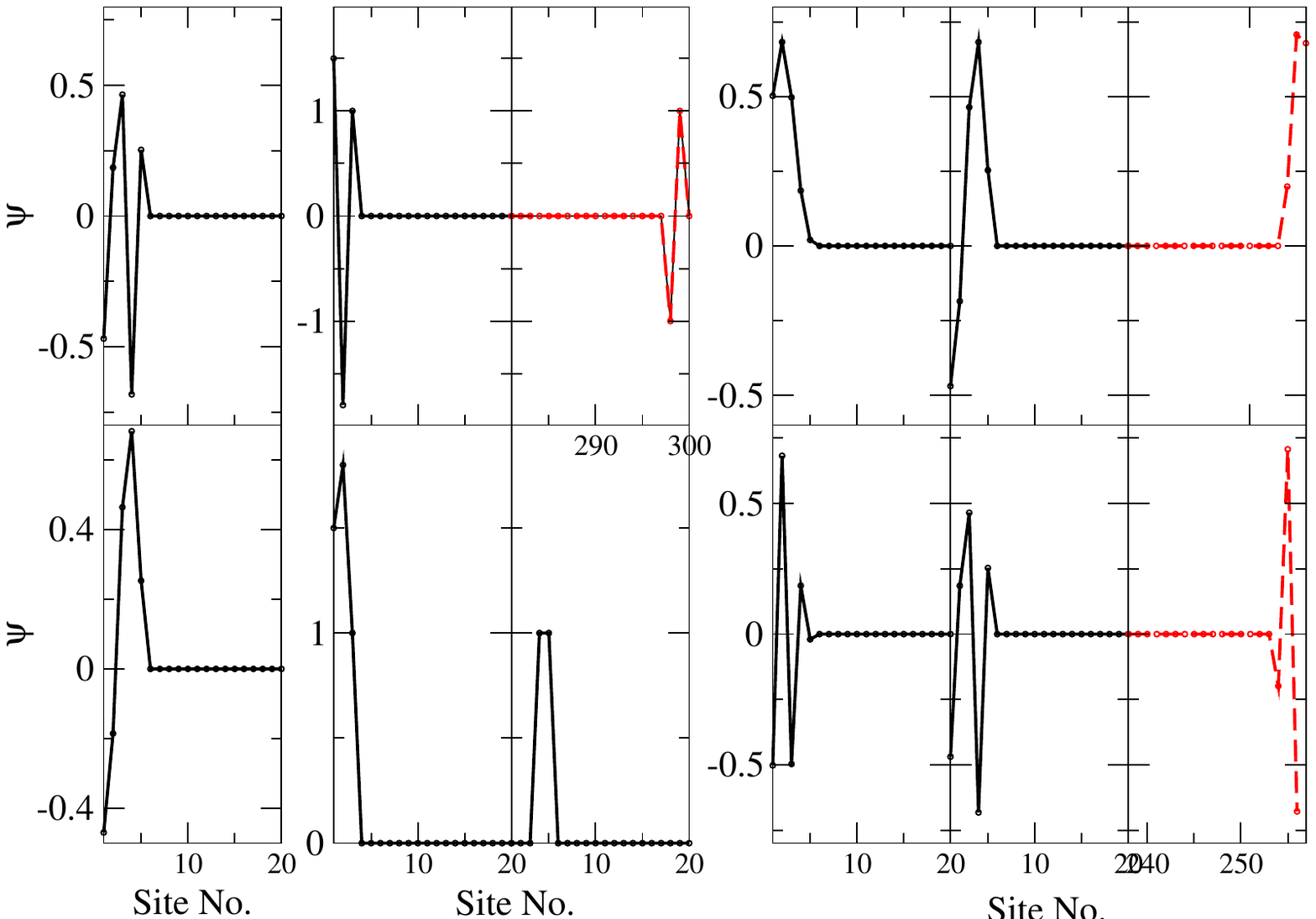}}
   \put(-40,140){(e)}
   \put(0,140){(f)}
   \put(80,140){(g)}
   \end{picture}
\caption{The mid-gap ZES of a finite SSH(like) chain for $\theta$ = (a) $\pi$, (b) $\pi/2$, (c) $\pi/3$ and (d) $\pi/4$. There are also in-gap end states at nonzero energies as can be seen for $\theta$ = (e) $\pi/2$, (f) $\pi/3$ and (g) $\pi/4$.}
\label{endstate}
\end{figure}

For the sake of completeness, we should mention here that a winding number of $\mathcal{W}=-1$ can be obtained in an extended SSH model incorporating longer range hopping between different sublattices\cite{winding}.
Interestingly, considering more than one further neighbor hopping term, P\'{e}rez-Gonz\'{a}lez \textit{et al.} (2018) demonstrated how an additional topological phase with $\mathcal{W}=2$ can be obtained\cite{gauss}. It shows two individual pairs of topological edge states. Moreover, the topological phases with larger winding numbers (say, up to $\mathcal{W}=4$) are also found to appear with the inclusion of multiple further neighbor hopping terms\cite{winding2}.

\subsection{End States for Different $\theta$'s}
Features of the end states for different $\theta$ were mentioned to some extent in Ref.\cite{skar}. Here we give some further details of the same for $\theta=\pi,~\pi/2,~\pi/3$ and $\pi/4$. Deep within the topological phase, the pair of chiral zero energy end states are obtained for $\theta=\pi$ that survive in single boundaries. However, close to QPT, the end states in NTP survive at both the boundaries and in the sublattices containing the end sites. The wave functions oscillate to gradually vanish within the bulk. They are of mixed chirality which can be combined linearly to yield chiral end modes surviving at single ends of the chain\cite{yang}. Similar trends are observed for other $\theta$ values as well. For {$\theta=2\pi/n$ with $n>2$, new in-gap states appear ($n-2$ in number, $i.e.$, 0, 2 and 6 for $n=$2, 4 and 8 respectively) in the NTP.} They can be called end states as they peak near the chain boundary (not necessarily at end sites though). However, we find only two in-gap states for {$n=6$ in the NTP}. Interestingly, there are few states {near the edge of these gaps} in the spectrum which show localized peaks near boundaries (see right panel plots of Fig.\ref{endstate}(f)).

\section{SSH(like) chain with domain wall}\label{sec4}
To embark on our exploration, we also introduce single domain wall\cite{domain} in our SSH(like) chain with the hopping modulation now taking the form of:
\begin{equation}\label{45}
\delta_{i}=d_{0}\tanh\Big[\frac{i-i_{0}}{\xi/a}\Big] \cos[(i-1)\theta],
\end{equation}
where $d_{0}$, $\xi$ and $a$ are the amplitude of the domain wall (DW), the width of the DW, and the lattice spacing respectively. Here, $i_{0}$ is a parameter that actually determines the position of the DW (in the present study, we consider the location of the DW at the center of the chain). This DW separates the two dimerized phases and the Hamiltonian (we call it $H_{ssh}^{dw}$) exhibits fractionalized zero modes at the location of the domain wall\cite{domain}. We can add here that the boundaries of the SSH(like) chain, we studied so far, may be seen as domain walls with the vacuum/surrounding.
\begin{figure}
   \vskip -.4 in
   \begin{picture}(100,100)
    \put(-70,0){
  \includegraphics[width=.5\linewidth]{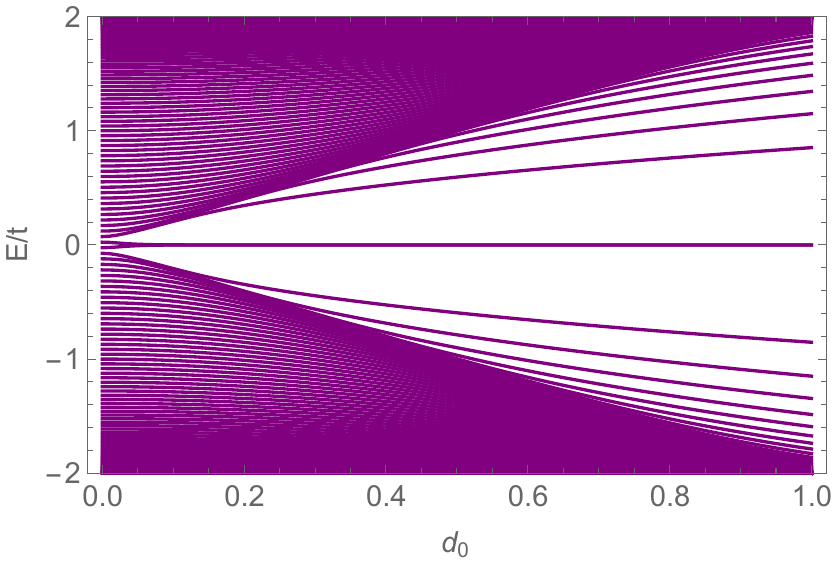}
  \includegraphics[width=.5\linewidth]{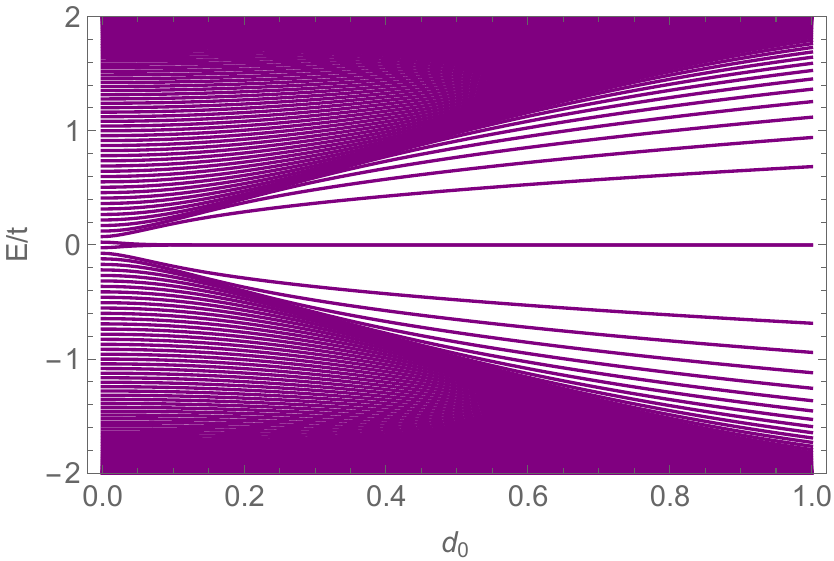}}
    \put(10,45){(a)}
    \put(120,45){(b)}
   \end{picture}\\
   \vskip -.2 in
   \begin{picture}(100,100)
     \put(-70,0){
   \includegraphics[width=.5\linewidth]{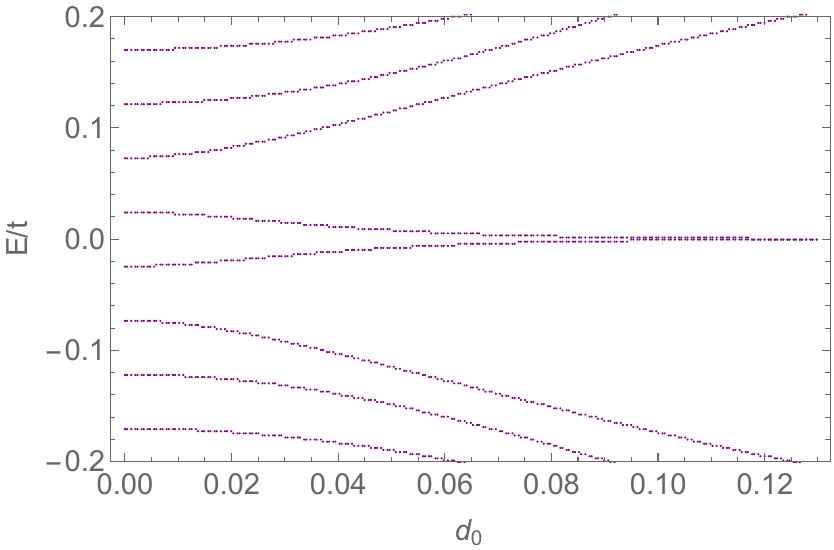}
   \includegraphics[width=.5\linewidth]{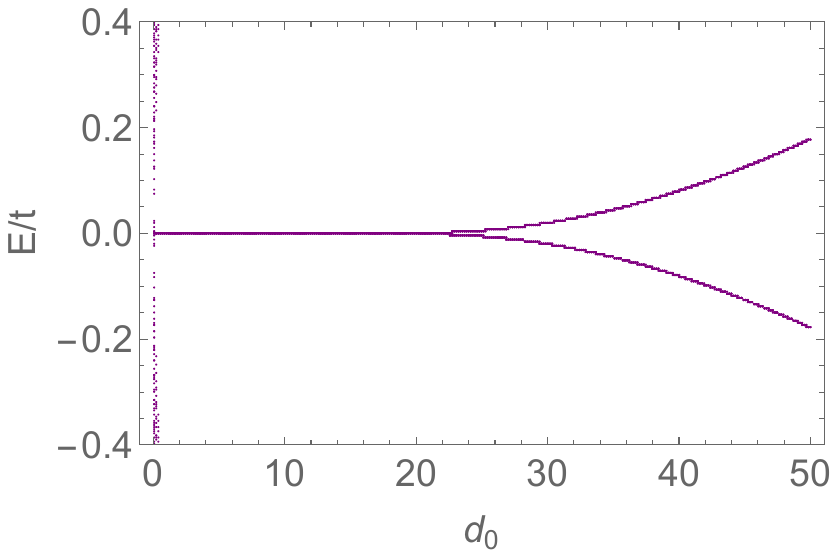}}
   \put(6,60){(c)}
     \put(148,60){(d)}
   \end{picture}  
\caption{Upper panels show the energy spectra of a SSH(like) chain with domain wall for $L = 128$ as a function of domain wall amplitude $d_{0}$ with $\theta=\pi$ for (a) $\xi/a=10$ and (b) $\xi/a=20$. The bottom panels depict the low energy spectra for (c) smaller and (d) higher variations of $d_{0}$ with $L=128$ and $\xi/a=20$.}
\label{fig6}
\end{figure}

\begin{figure}
   \vskip -.4 in
   \begin{picture}(100,100)
     \put(-70,0){
  \includegraphics[width=.5\linewidth]{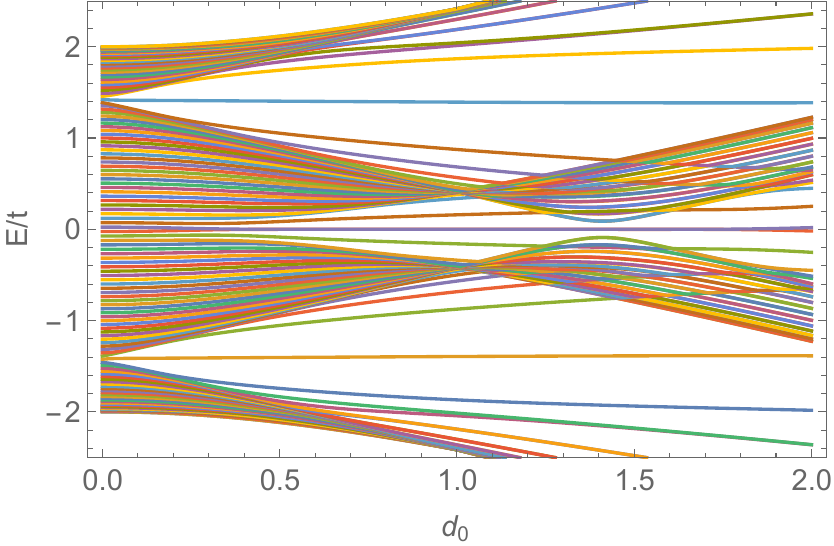}
  \includegraphics[width=.5\linewidth]{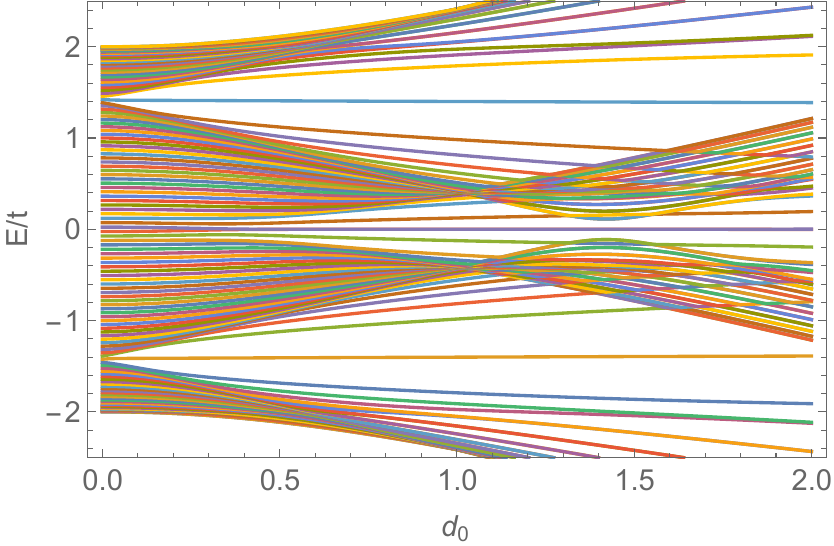}}
     \put(20,62){(a)}
     \put(120,62){(b)}
   \end{picture}\\
   \vskip -.2 in
   \begin{picture}(100,100)
     \put(-70,0){
   \includegraphics[width=.5\linewidth]{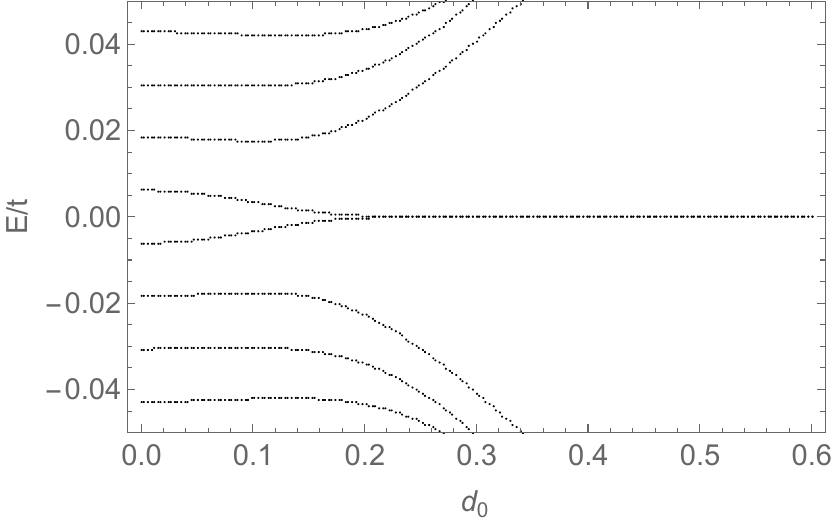}
   \includegraphics[width=.5\linewidth]{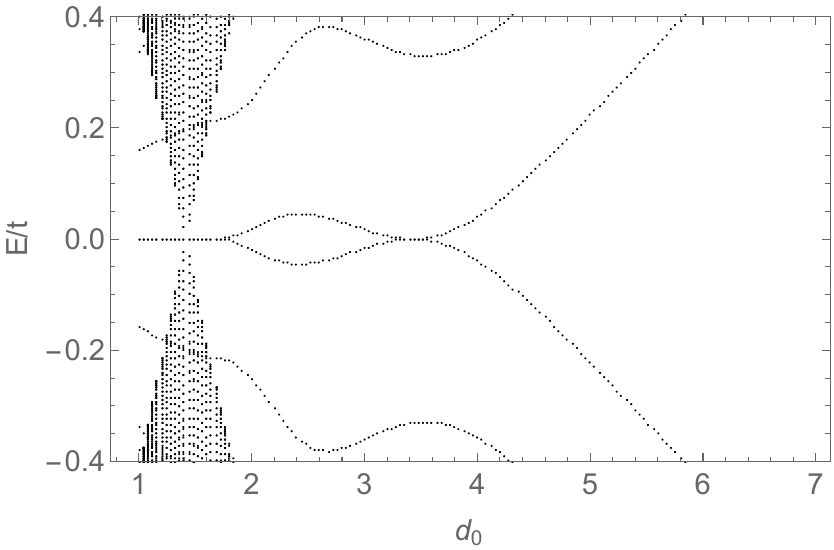}}
   \put(30,58){(c)}
     \put(110,60){(d)}
   \end{picture}  
\caption{Top panels present the energy spectra of a SSH(like) chain with domain wall for $L = 128$ as a function of domain wall amplitude $d_{0}$ for $\theta=\pi/2$ with (a) $\xi/a=10$ and (b) $\xi/a=16$ respectively. The bottom panels show the corresponding low energy states for (c) smaller and (d) higher $d_{0}$ for a chain with $L=512$ and $\xi/a=10$.}
\label{fig7}
\end{figure}

\begin{figure}
   \vskip -.4 in
   \begin{picture}(100,100)
     \put(-70,0){
  \includegraphics[width=.5\linewidth]{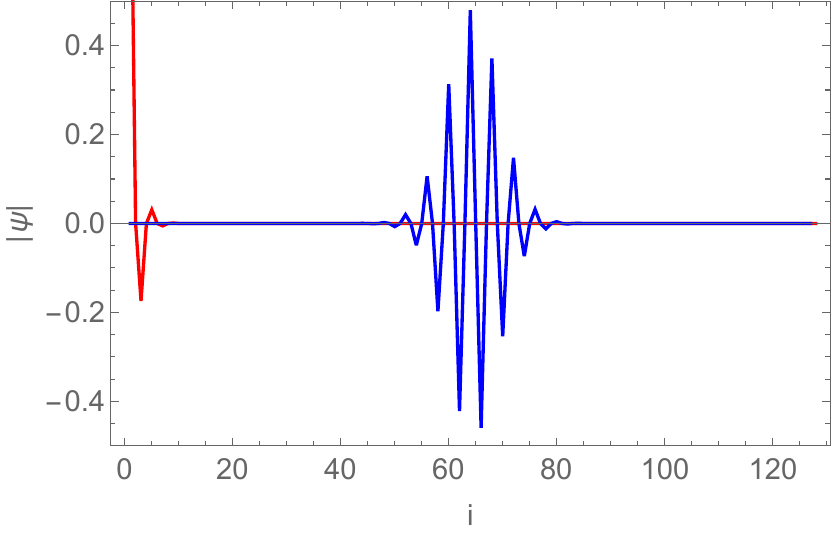}
  \includegraphics[width=.5\linewidth]{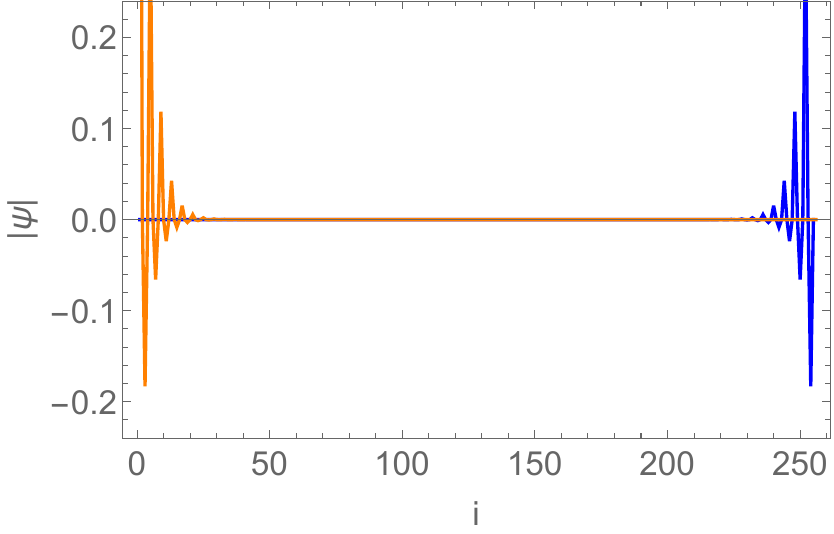}}
     \put(-28,60){(a)}
     \put(110,60){(b)}
   \end{picture}\\
   \vskip -.2 in
   \begin{picture}(100,100)
     \put(-70,0){
   \includegraphics[width=.5\linewidth]{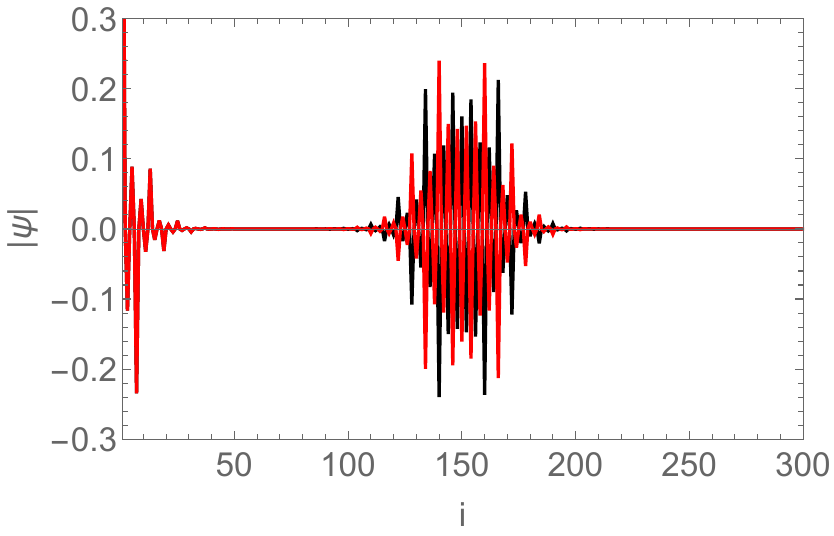}
   \includegraphics[width=.5\linewidth]{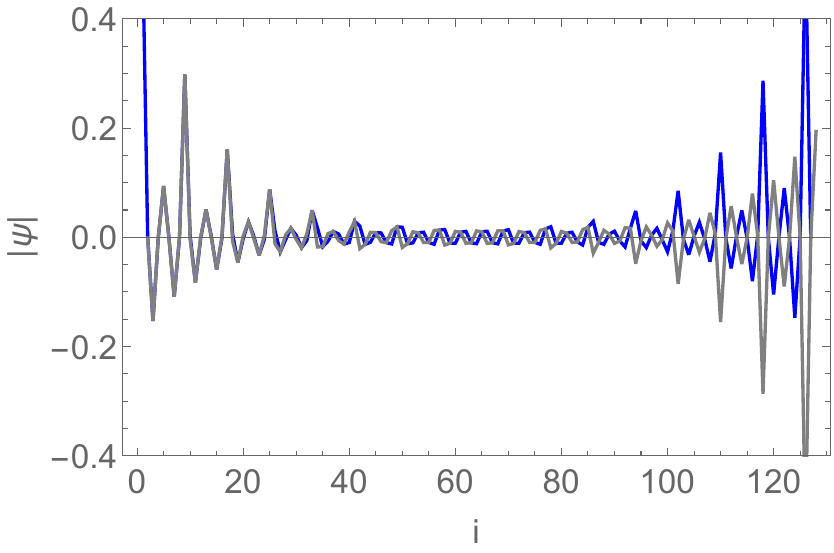}}
   \put(-28,60){(c)}
     \put(110,60){(d)}
   \end{picture}  
   \caption{Zero energy states of SSH(like) chain in presence of domain wall and for choice of parameters (a) $\theta=~\pi$, $d_{0}=~0.7$, $\xi/a=~16$, $L=~128$ (b) $\theta=~\pi/2$, $d_{0}=~0.7$, $\xi/a=~16$, $L=~256$, (c) $\theta=~\pi/3$, $d_{0}=~0.9$, $\xi/a=~16$, $L=~300$ and (d) $\theta=~\pi/4$, $d_{0}=~0.9$, $\xi/a=~16$, $L=~128$.}
\label{fig6a}
\end{figure}

As part of our study, we probe such systems with a single static domain wall for different values of $\theta$. In the sequel, the energy spectra of a SSH(like) chain in the presence of a domain wall as a function of domain wall amplitude $d_{0}$ for $\theta=\pi$ and different values of $\xi$ is presented in Fig.\ref{fig6}. We find the DW results in additional in-gap states {(often called bound states\cite{domain})} other than the mid-gap ZES, and the gap reduces with an increase of $\xi/a$. Interestingly, the zero modes disappear for smaller and higher values of amplitude ($d_{0}$). {One gets a nearest neighbor (NN) tight binding model at $d_0=0$ where no zero energy state is there in a finite system. For small $d_0$ on the other hand, the availability of topologically protected zero modes depends on the length of the chain. For a longer chain one gets a smaller $d_0$ cut-off that denies the existence of the zero mode. The ZES vanishes for very large $d_0$ as well. In fact, the decay of the end states away from the edges goes as $\sim|\frac{t-d_0}{t+d_0}|$\cite{skar}, which indicates a slow decay for both small and large $d_0$ values. These lead to hybridization of the two degenerate end modes producing symmetric and antisymmetric linear combinations that lie (slightly) above and below the zero energy.} The presence of a pair $(E,-E)$ in the energy spectra (see Fig.\ref{fig6} and Fig.\ref{fig7}) ensures the persistence of chirality even after the introduction of DW\cite{sang}.
The wave function for zero modes for this case is shown in Fig.\ref{fig6a}(a). The figure illustrates that one, among two zero modes, is localized at the DW position and the other at one end of the chain (on the different sublattice than where the domain wall resides). This result implies that the domain walls, similar to the boundary of the SSH chain, host zero energy localized states. However, the scenario changes for other values of $\theta$.

For $\theta=\pi/2$, there also appear additional in-gap states including topological zero modes, and is presented in Fig.\ref{fig7}. Like in $\theta=\pi$, ZES will no longer be present for small and very large values of $d_{0}$. More interestingly, for intermediate $d_{0}$ values, we find the zero modes to reappear within a small window of $d_0$ values [see Fig.\ref{fig7}(d)].

In Fig.\ref{fig6a}, we show typical ZES for $\theta=\pi,~\pi/2,~\pi/3,~\pi/4$.
We find the localized states to appear at DW position only for $\theta=\pi,~\pi/3$ values whereas for $\theta=\pi/2,~\pi/4$ they appear only at the boundaries. In general, these solitonic states appear for $\theta=\frac{\pi}{2p+1}$ with $p$ taking all integer values and they carry a fractional charge of $\pm e/2$ when electrons are spinless\cite{kane,yang}. The $\pm e/2$ charged state appears when typical ZES is empty/occupied. In the strong coupling limit ($|\Delta|=t$), there will be an unpaired state on the DW at zero energy. They persist with the reduction of $\Delta$ but perish at $\Delta\rightarrow0$. Such typical ZES are sometimes known as domain wall solitons (kink) and are the physical realization of ZES found in a one-dimensional field theory proposed by Jackiw and Rebbi (JR)\cite{JR}. 

The SSH model portrays free fermions on a lattice with the hopping strengths alternating between weak and strong bonds and the continuum limit of this model delineates a massive
Dirac fermion\cite{sshjk}. On the other hand, the JR model\cite{JR} can be viewed as a one-dimensional system where Dirac fermions can be coupled to a soliton field. One can introduce a topological defect (domain wall) familiar as a soliton in the former by changing the arrangements of the hopping strengths (e.g. weak-strong to strong-weak) at a certain lattice site while in the latter by tuning the mass of
the fermion such that it changes sign at a certain point. These solitons mark an interface between a non-topological and a topological phase and accordingly, they have several notable features. For instance, they host localized ZES, which are characteristic of a transition from non-topological to topological phases and vice-versa, and they also hold fractional charges\cite{JF}. In the continuum limit, these soliton induced charge fractionalizations can be ascribed to the local charge operators showing fractional eigenvalues rather than just having fractional expectation values\cite{jackiw}. The ZES can be called excitations on top of this fractionally charged background and they can be related to MZMs\cite{ryland}.

This soliton appears due to the presence of a single DW. However, the introduction of another DW may lead to the existence of an anti-soliton resulting in a soliton and anti-soliton pair in the chain. Interestingly, in contrast to
polyacetylene, a single DW in graphene nanoribbons can support both a soliton and anti-soliton (for more details see Ref.\cite{yang}).

{Fascinatingly, the engagement of the DW can also show sharp peaks at the DW position for the in-gap and bound states.}

\section{Disorder}\label{sec4a}

Nowadays, it has become very important to investigate the effect of disorder in electronic systems. Several recent studies have included the behavior of topological systems considering random-dimer disorder\cite{dimer1,dimer2}, quasi-periodic disorder \cite{quasi}, and strong disorder\cite{strong1,domain}.

This section aims to study numerically the effect of on-site disorder (diagonal) and hopping disorder (off-diagonal) on the chirality and zero energy states of the SSH(like) chain. We introduce on-site disorder via
\begin{equation}\label{46}
H_{on}^{disorder}=\sum_{i=1}^{L}\epsilon_{i}c_{i}^{\dagger}c_{i},
\end{equation}
where $\epsilon_{i}$ can be a constant, random (taken from a uniform distribution on $[-G,~G]$\cite{domain,strong1}), staggered (respecting the original 2 sublattice structure of SSH chain) or interpolated potential. In Ref.\cite{gauss}, authors have considered the on-site energies as the random numbers taken from a Gaussian distribution centered at zero. However, the on-site disorder considered herein is exceedingly similar to the strong disorder mentioned in Refs.\cite{domain,strong1}. We find that the nontriviality of the spectrum and states obtained due to the effect of disorder on the SSH(like) chain with a multi-sublattice structure, is worth mentioning and is distinct from the pioneer studies\cite{dimer1,dimer2,quasi,strong1}. The inclusion of the random potential disrupts the chiral symmetry resulting in the disappearance of fractionalization modes observed at the clean limit ( $i.e.,~G=0$ or $\epsilon_{i}=0$)\cite{domain}. Interestingly, however, the end states can retain their chiral nature even for strong (random) on-site disorder due to the effect of Anderson localization\cite{domain,anderson,ander}.

\begin{figure}[t]
   \vskip -.4 in
   \begin{picture}(100,100)
     \put(-70,0){
  \includegraphics[width=.5\linewidth]{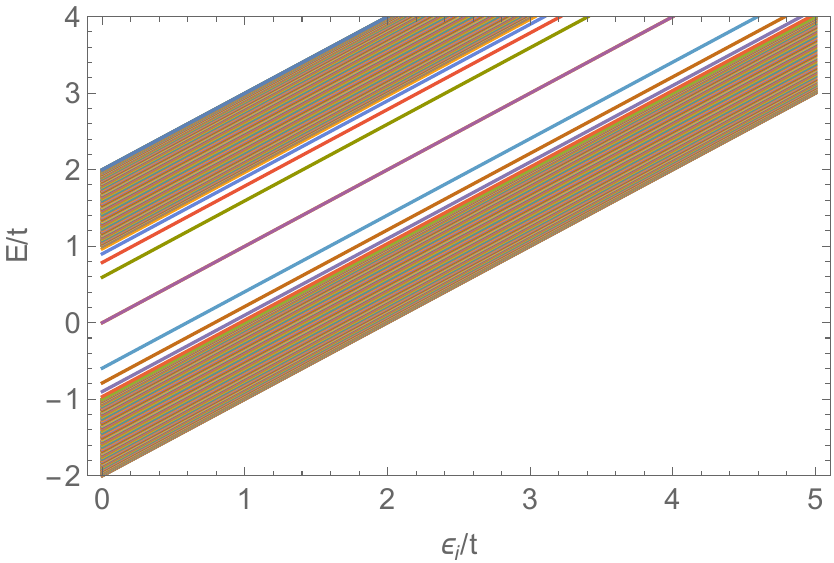}
  \includegraphics[width=.5\linewidth]{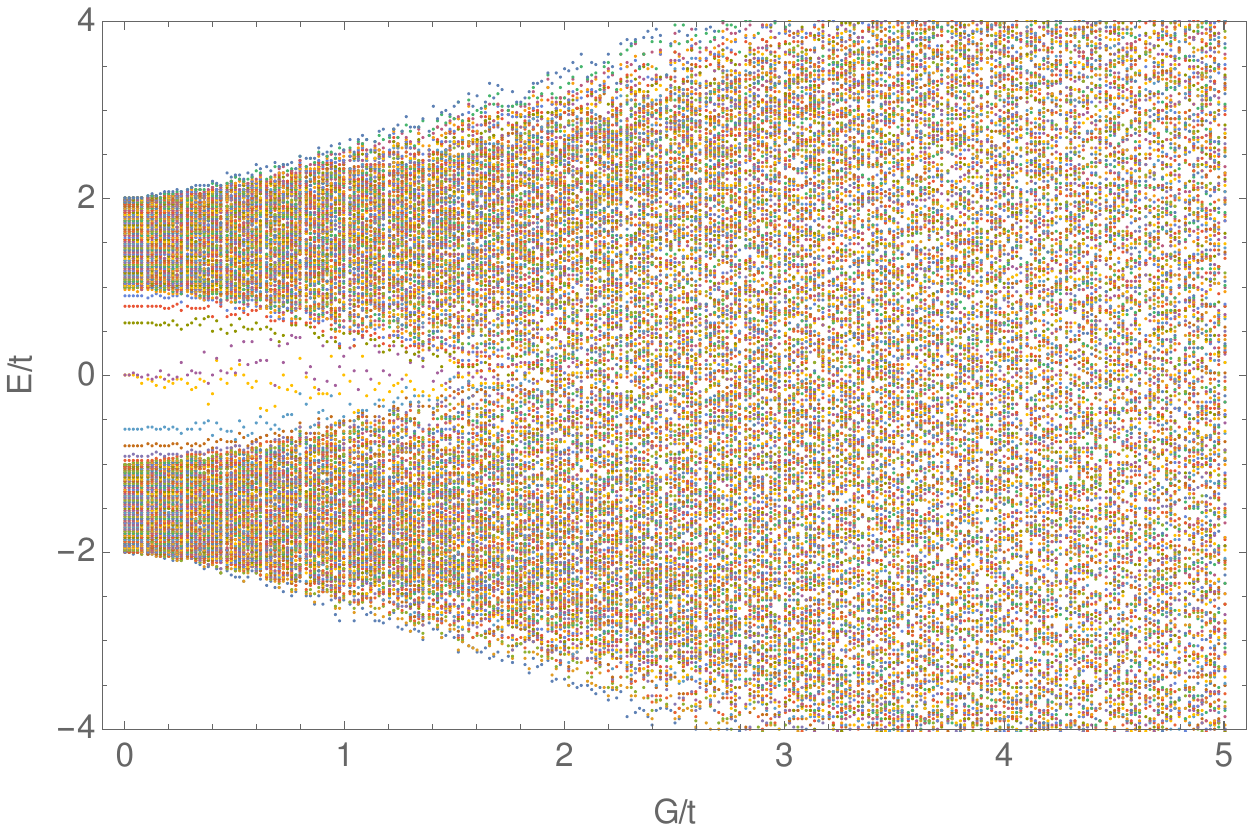}}
     \put(-49,68){(a)}
     \put(70,68){(b)}
   \end{picture}\\
   \vskip -.2 in
   \begin{picture}(100,100)
     \put(-70,0){
   \includegraphics[width=.5\linewidth]{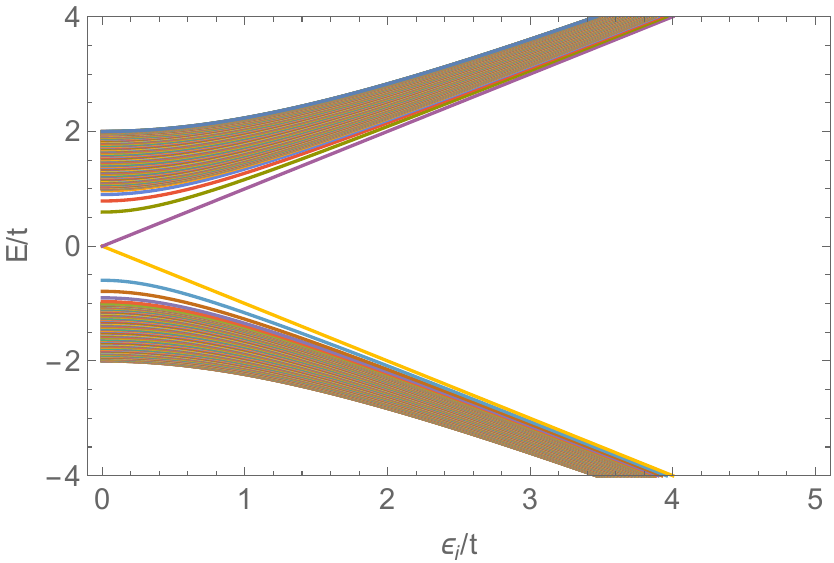}
   \includegraphics[width=.5\linewidth]{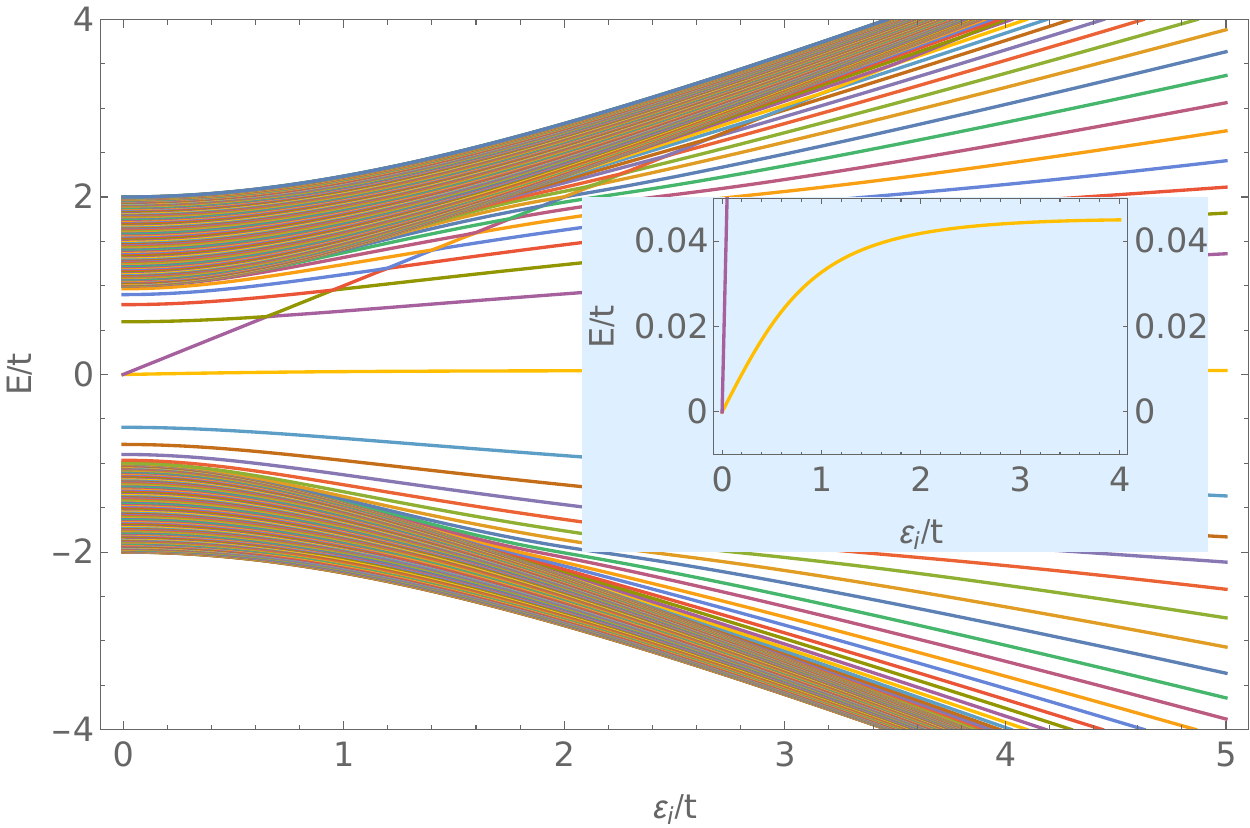}}
   \put(-49,70){(c)}
     \put(70,70){(d)}
   \end{picture}  
\caption{The energy spectrum of a SSH(like) chain in presence of various onsite disorders ((a) constant, (b) random, (c) RM type, (d) AI type) for $\theta=\pi$ and a single DW at the center as a function of disorder strength $\epsilon_{i}/t$ or $G/t$. Other choices of parameters are $L=~256$, $\xi/a=~10$, $d_{0}=~0.5$. The inset in (d) shows the low energy states.}
\label{fig9}
\end{figure}

\begin{figure}[t]
   \vskip -.4 in
   \begin{picture}(100,100)
     \put(-70,0){
  \includegraphics[width=.5\linewidth]{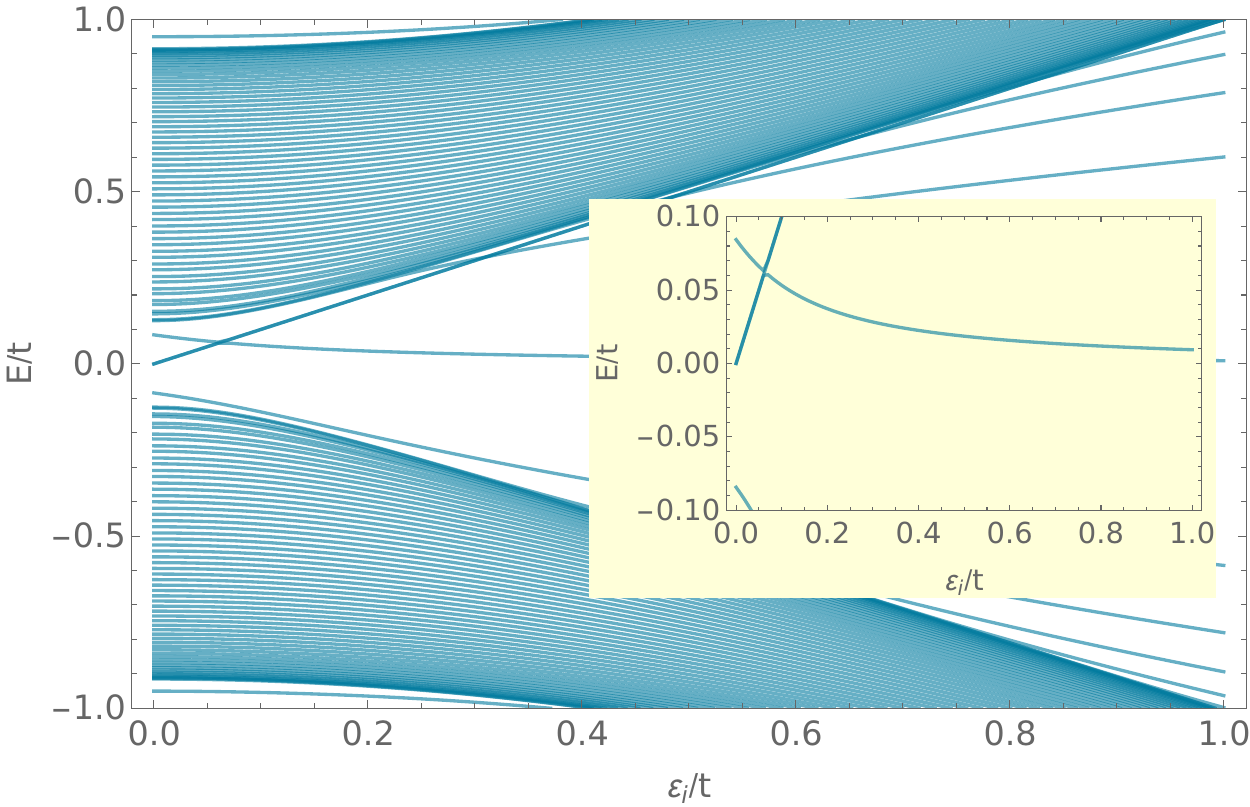}
  \includegraphics[width=.5\linewidth]{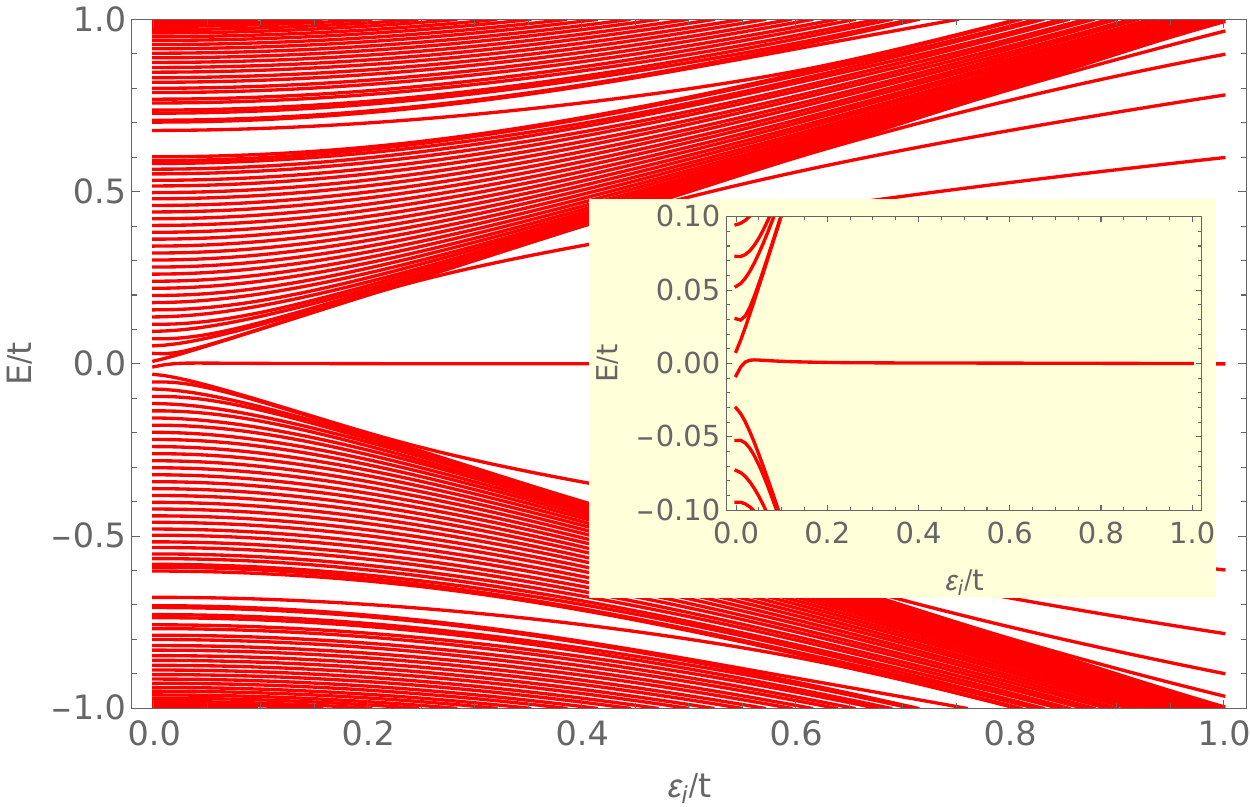}}
     \put(30,60){(a)}
     \put(150,60){(b)}
   \end{picture}  
\caption{The ordered energy eigenvalues of a SSH(like) chain in the presence of an AI-type DW/disorder for (a) $\theta=\pi/2$ and (b) $\theta=\pi/4$ as a function of disorder strength $\epsilon_{i}/t$  with the same choice of parameter as in Fig.\ref{fig9}. The insets show the low energy states.}
\label{fig10}
\end{figure}

In this regard, it is worth mentioning of RM DW or that of AI class which consider staggered or interpolated on-site potentials as disorder respectively\cite{onsite,sang}. While SSH and AI domain walls support chiral symmetry protected ZES, the RM configuration features DW states at nonzero energies\cite{onsite,sang}. This section is thus devoted to investigating how the chirality-preserving bound states of SSH(like) chain evolve with disorder of different kinds ($i.e.,$ constant, random, staggered or interpolated $\epsilon_i$) in the presence of the periodic hopping modulations. How the zero modes dissipate their chirality with an increasing disorder strength ($G/t$ or $\epsilon_{i}/t$) can be understood from the spectrum of the Hamiltonian $H=H_{ssh}^{dw}+H_{on}^{disorder}$ and is reported in Figs.\ref{fig9}-\ref{fig10}. The contravention of chiral symmetry affects the energy spectrum which is no longer symmetric i.e. for every eigenstate having energy $E$, its chiral symmetric partner with energy $-E$ does not necessarily exist.

We find for $\theta=\pi$ and for $L=256$ and $\xi/a=10$, the spectrum is legibly divided into two bands with disorder strength (see Fig.\ref{fig9}). The band separation increases (decreases) with $\epsilon_{i}/t$ for RM and AI type (random) disorder and the ZES vanishes for nonzero disorder strength. However, for AI type disorder a DW state is observed close to zero energy {while a random disorder features fluctuating zero modes\cite{domain}.} Apart from ZES, several other in-gap states also appear between the two bulk bands for different choices of parameters (such as $d_{0}$, $\xi/a$, and $L$). In particular, increases in DW amplitude and $\xi/a$, in turn, give rise to an increase in the number of in-gap states. An infinite SSH chain can have unpaired zero-energy states but a finite chain with open boundaries always leads to paired zero-energy states\cite{ns}. We can mention here of a related model called Shiozaki-Sato-Gomi or SSG model that features constant NN hopping but staggered onsite potentials and obeys a non-symmorphic (NS) chiral symmetry. In contrast to the SSH model, this model supports unpaired zero-energy states at the position of smooth DW and not at sharp interfaces with vacuum (see Ref.\cite{ns}). 

For $\theta=\pi/2$, $\theta=\pi/3$ or $\pi/4$, we find that the {periodically hopping modulated} SSH model with AI DW/disorder retrieves the ZES with an increase in disorder strength $\epsilon_{i}/t$ (see Fig.\ref{fig10}). 
These ZES are located at the position of the DW. For random type disorder, the energy gap closes with the increase of {$G/t$}. The number of in-gap states as well as the critical disorder strength of the closure of the energy gap decreases as $\theta$ is reduced in the commensurate manner as considered in this paper. Considering the commensurate
$\theta$ values and taking into account the various onsite disorders, one can see that there is no such
remarkable change in the energy spectrum.

Other than the onsite potential, disorder can come as hopping parameters as well. The study of the chiral-symmetry-preserving hopping disorder via the Hamiltonian
\begin{equation}\label{47}
H_{hop}^{disorder}=\sum_{i=1}^{L-1}\tau_{i}(c_{i+1}^{\dagger}c_{i}+c_{i}^{\dagger}c_{i+1}),
\end{equation}
with $\tau_{i}$ randomly generated from a uniform distribution, have shown polarization, as a real space estimator of topological invariance, to continuously reduce to zero with increasing disorder strength for both edge and domain wall states, though their localization remains intact for all disorder strengths\cite{domain}. Unlike the on-site disorder, the hopping disorder respects the chiral symmetry. Consequently, the robustness of the DW/ edge states are expected in the presence of this kind of disorder. However, notice that the Hamiltonian $H_{hop}^{disorder}$ merely renormalizes the hopping parameters of $H_{ssh}^{dw}$.

This type of perturbation preserves sublattice symmetry which in turn gives non-fluctuating zero energy modes. A strong hopping disorder, however, destroys
their chirality resulting in fluctuating zero energy modes\cite{domain,strong1}. Here, we mention of the related AIII chiral-symmetric system which breaks its chiral symmetry at critical disorder strength resulting
in abrupt changes in winding number from 1 to 0 \cite{strong1}. Moreover, this abrupt change in winding number in the presence of strong hopping disorder for SSH chain with periodic modulated hopping is worth studying and we will leave this issue for future communication.

\section{Summary}\label{sec5}

In this paper, we have tried to accumulate important findings on SSH model spectra, topology as well as the mid-gap zero energy states in the absence as well as the presence of DW and disorder and at the same time, adding new insights on the same while introducing additional periodic modulation in the hopping parameter. Firstly without artificial domain walls and disorder, we find that for hopping modulation with commensurate frequency, new in-gap end-states appear - more for smaller values of it. The topological regime shows interesting variation with $\theta$. We find one, two, three, and four topological phase transition points for $\theta=\pi,~\pi/2,~\pi/3$, and $\pi/4$ respectively. In the presence of a static DW that has been put artificially at the center of a finite chain, one end state gets depleted while one ZES appears at the position of the DW for $\theta=\frac{\pi}{2p+1}$ for zero or an integer $p$ value. Unlike the SSH type DW, an AI type DW for the model doesn't support chiral symmetry protected ZES while an RM DW shows {both the edge and} domain wall states to be at nonzero energies. However, an onsite disorder always disrupts the system's chirality. With the commensurate reduction of $\theta$, more in-gap states appear and the ZES of a clean system moves to nonzero energies, though for AI type disorder the DW state reaches close to zero energy with disorder strength. A hopping disorder, on the other hand, doesn't contribute much new physics in this regard as a weak hopping disorder still respects the chiral symmetry.

One can verify these results of hopping modulated SSH(like) chains in cold atom systems within optical lattices\cite{xie}, or maybe in specially designed graphene nanoribbons\cite{gronning} or topological acoustic systems\cite{yang2}. Experimental confirmation of similar outcomes can lead to further manipulation of these periodically modulated hopping models to look for more exotic behavior. Down the line, we also have plans to study the out of equilibrium behavior of such extended SSH model, subjected to a quantum quench which has shown to lead to an effective metal insulator transition for $\theta=\pi$\cite{porta}. For a high frequency periodic quench, it will also be interesting to do a Floquet analysis\cite{floquet} and probe the competition between the topology and the time periodic driving in our SSH(like) models.

\section*{Acknowledgements}
SK thanks G. Baskaran, H. Yao, B. Kumar, S. Basu, S. Mandal and A. Saha for fruitful discussions. This work is financially supported by DST-SERB, Government of India via grant no. CRG/2022/002781.
\section*{Appendix~A: Winding Number in Terms of Poles and Zeros}
The winding number can be defined in terms of poles and zeros of ${\bf Det[V(k)]}=f(k)$ as a function of a complex variable $z(k)=e^{-ik}$ inside the unit circle. Accordingly, the winding number becomes\cite{s5d}
\begin{equation}\label{30a}
\mathcal{W}=\frac{1}{2\pi i}\oint_{|z|=1} \frac{f'(z)}{f(z)}dz
\tag{A1}\end{equation}
Here, in the complex plane, the closed curve represents the anti-clockwise path around the unit circle. Considering the number of zeros ($U_{f}$) and number of poles ($V_{f}$) of $f(z)$ enclosed by the curve, one can write, following Cauchy's argument principle of complex analysis\cite{ole,s5e}, 
\begin{equation}\label{30b}
\mathcal{W}=U_{f}-V_{f}
\tag{A2}\end{equation}
Every zero and pole are weighted by their multiplicity\cite{cmnt2} and order respectively. It is presumed that there are no zeros and poles on the curve. So for this present case of $\theta=\pi/2$, we get $f(z)=(t^2-\Delta^2)+t^2z$. $f(z)$ contains no poles, but it has a zero of multiplicity one:
$$f[-(t^2-\Delta^2)/t^2]=f(z_{0})=0$$
The zero is inside the unit circle if $|t^2-\Delta^2|<|t^2|$. Therefore, Eq.(\ref{30b}) gives 
\begin{equation}
{\mathcal{W}=U_{f}-V_{f}=\Bigg\{\begin{matrix}
    1-0=1, & 0<\Delta^2/t^2<2; \\
    0, & \Delta^2/t^2>2;\\
    \text{undefined}, &\Delta/t=0
  \end{matrix}}
\tag{A3}\end{equation}
which is the same as Eq.(\ref{9a}). If {$\Delta/t=0$} then the spectrum is considered to be conducting and the representation of topological invariants is not well-defined. For this case, contrary to the above conjecture, the zero of $f(z)$ arises at $z_{0}=-1$ which is on the unit circle.
\section*{Appendix~B: Berry Phase}
Another topological invariant of this $1D$ system is the Zak phase. The bulk-boundary correspondence affirms that the presence of edge states is related to the nonzero topological invariant of the bulk. In particular, the number of edge modes on every edge is exactly equal to $|\mathcal{W}|$. The Zak phase $\gamma$\cite{s5b,s5c,s7} of the system corresponds to $\pi\mathcal{W}$ (eventually, depending on the convention, modulo 2$\pi$).  Closing of the energy gap (for {$\Delta/t=0$}, in this case), and a subsequent re-opening at the Brillouin zone boundaries indicates that the system may go through a topological phase transition. In addition, the appearance of a nonzero topological invariant in one of the gapped phases ensures the advent of nontrivial topology. This topological invariant is expressed in terms of the Zak phase which is purely a bulk property of the system. Therefore, we need to make sure of the accomplishment of the Born-von Karman periodic boundary condition.

It is known that the 1D winding number in the SSH model is closely related to the Zak phase\cite{s7}, which is basically similar to the Berry phase ($\gamma$)\cite{s8} for 1D systems. Consequently, it is related to the winding number as $\textbf{$\gamma$}=\pi \mathcal{W}$. So, the Berry phase becomes
\begin{equation}\label{berrya}
{\mathbf{\gamma} = \Bigg\{\begin{matrix}
     \pi, & 0<\Delta^2/t^2<2; \\
    0, & \Delta^2/t^2>2;\\
    \text{undefined}, & \Delta/t=0
  \end{matrix}}
\tag{B1}\end{equation}

\section*{Appendix~C}
The Bloch Hamiltonian in the case of $\theta=\pi/3$ is given in Eq.(\ref{32cc}) in which $z=e^{-6ik}$. Here, the unit cell contains six sublattices, so the size of the BZ boundary further reduces i.e., $k\in[-\pi/6,~\pi/6$].
\begin{align}\label{32cc}
H_{k} =
\begin{pmatrix}
0 & (t+\Delta) & 0 & 0 & 0 & (t+\frac{\Delta}{2})z \\
(t+\Delta) & 0 & (t+\frac{\Delta}{2}) & 0 & 0 & 0 \\
0 & (t+\frac{\Delta}{2}) & 0 & (t-\frac{\Delta}{2}) & 0 & 0 \\
0 & 0 & (t-\frac{\Delta}{2}) & 0 & (t-\Delta) & 0 \\
0 & 0 & 0 & (t-\Delta) & 0 & (t-\frac{\Delta}{2}) \\
(t+\frac{\Delta}{2})z^* & 0 & 0 & 0 & (t-\frac{\Delta}{2}) & 0 \\
\end{pmatrix}
\tag{C1}\end{align}

The Bloch Hamiltonian for $\theta=\pi/4$ is calculated as in Eq.(\ref{32d}). Here, eight sublattices are embedded in a unit cell which can further partition the BZ boundary. Thus in reduced BZ, $k\in[-\pi/8,~\pi/8$].
\begin{widetext}
\begin{equation}\label{32d}
H_{k} =
\begin{pmatrix}
0 & (t+\Delta) & 0 & 0 & 0 & 0 & 0 & (t+\frac{\Delta}{\sqrt{2}})e^{-8ik} \\
(t+\Delta) & 0 & (t+\frac{\Delta}{\sqrt{2}}) & 0 & 0 & 0 & 0 & 0 \\
0 & (t+\frac{\Delta}{\sqrt{2}}) & 0 & t & 0 & 0 & 0 & 0 \\
0 & 0 & t & 0 & (t-\frac{\Delta}{\sqrt{2}}) & 0 & 0 & 0 \\
0 & 0 & 0 & (t-\frac{\Delta}{\sqrt{2}}) & 0 & (t-\Delta) & 0 & 0\\
0 & 0 & 0 & 0 & (t-\Delta) & 0 & (t-\frac{\Delta}{\sqrt{2}} & 0 \\
0 & 0 & 0 & 0 & 0 & (t-\frac{\Delta}{\sqrt{2}}) & 0 & t \\
(t+\frac{\Delta}{\sqrt{2}})e^{8ik} & 0 & 0 & 0 & 0 & 0 & t & 0 \\
\end{pmatrix}
\tag{C2}\end{equation}
\end{widetext}


\begin{thebibliography}{00}
\bibitem{cmntMF} Ref.\cite{skar} is over-ambitious in calling the zero energy modes of SSH(like) chain as Majorana modes. In fact, a pair of Majorana modes of different flavors localized at each end is superposed to yield electronic end modes there.
\bibitem{cmnt-chiral}  We should note here that the first quantized single particle chiral symmetric Hamiltonian reverses sign under the chiral symmetry operation (see also Sec. IIA1). This sign change is however compensated by the anti-commutivity of the fermionic operators to produce the invariance of the second quantized Hamiltonian\cite{ludwig,ole}. A Fourier transformed 2nd quantized SSH Hamiltonian, written as $H=\sum (a_k^\dagger~b_k^\dagger)H_k(a_k~b_k)^T$ as in Ref.\cite{skar}, reserves its sign under the 2nd quantized chiral symmetry transformation given as $a_k(b_k)\rightarrow b_k^\dagger(a_k^\dagger),~H_k\rightarrow -H_k$.
\bibitem{comment} Even though the hopping strength is complex, one can consider a gauge transformation $c_{j}\rightarrow e^{i\alpha j}c_{j}$ to get the hopping amplitudes real again. In other words, even with complex hopping strength, the model retains time-reversal symmetry by defining the correct time-reversal operator.
\bibitem{cmnt2} A zero in the form $a_{x}(z-z_{0})^{x}$ has a multiplicity $x$ and a pole in the form $a_{y}=\frac{1}{(z-z_{0})^{y}}$ has a order $y$.
\end{thebibliography}
\end{document}